\documentclass[12pt]{article}
\pdfoutput=1
\usepackage{jcappub}

\usepackage{url}
\usepackage{physics}

\usepackage{multirow}
\usepackage{booktabs}
\usepackage{placeins}

\newcommand{\factor}[3]{\left(\frac{#1}{#2\unit{#3}}\right)}
\newcommand{\unit}[1]{\,\mathrm{#1}}

\newcommand{\iu}{\mathrm{i}}
\newcommand{\me}{\mathrm{e}}
\newcommand{\micro}{\mu}

\renewcommand{\vec}[1]{\boldsymbol{#1}}
\renewcommand{\d}{\mathrm{d}}

\def\epsilon{\varepsilon}
\def\theta{\vartheta}
\def\lsim{\raise0.3ex\hbox{$\;<$\kern-0.75em\raise-1.1ex\hbox{$\sim\;$}}}
\def\gsim{\raise0.3ex\hbox{$\;>$\kern-0.75em\raise-1.1ex\hbox{$\sim\;$}}}

\newcommand{\apj}{{Astrophys.\ J. }}
\newcommand{\apjl}{{Astrophys.\ J.\ Lett. }}

\title{On the origin and the detection of characteristic axion wiggles in
photon spectra}

\author{M. Kachelrie{\ss} and}
\author{J. Tjemsland}

\affiliation{Institutt for fysikk, NTNU, Trondheim, Norway}

\keywords{axion, axion-like particles, photon-ALP oscillation, 
turbulent magnetic fields}

\abstract{
Photons propagating in an external magnetic field may oscillate into
axions or axion-like particles (ALPs). Such oscillations will lead to
characteristic features in the energy spectrum of high-energy photons from
astrophysical sources that can be used to probe the existence of ALPs.
In this work, we revisit
the signatures of these oscillations and stress the importance of a proper
treatment of turbulent magnetic fields. 
We implement axions into {\tt ELMAG},
a standard tool for modelling in a Monte Carlo framework the propagation
of gamma-rays in the Universe, complementing thereby
the usual description of photon-axion oscillations with a Monte Carlo
treatment of high-energy photon propagation and interactions.
We also propose an alternative method of detecting axions through
the discrete power spectrum using as observable
the energy dependence of wiggles in the photon spectra.
}

\begin{document}

\maketitle


\section{Introduction}

The Standard Model of particle physics has attained immense success over the
years. Yet, it has several theoretical shortcomings such as 
the strong CP problem. Moreover, there are some experimental indications
for its incompleteness, with the absence of a suitable dark
matter candidate as the most striking one. As a solution to the
strong CP problem, Peccei and Quinn~\cite{Peccei:1977hh,Peccei:1977ur}
postulated in 1977 the existence of an additional U(1) symmetry that is
spontaneously broken, thereby giving rise to a Nambu-Goldstone boson---the
axion $a$. The two-gluon-axion vertex introduced to solve the strong CP
problem induces a small axion mass through pion mixing,
$m_af_a\approx m_\pi f_\pi$, degrading the axion to a pseudo-Goldstone
boson~\cite{Weinberg:1977ma,Wilczek:1977pj}. Intriguingly, this promotes
the axion into a suitable cold dark matter candidate
despite its small mass if it is produced through, e.g., the
so-called misalignment mechanism~\cite{Preskill:1982cy,Abbott:1982af,Dine:1982ah,Marsh:2015xka}.
Other light pseudo-scalars bosons which have the same
characteristic two-photon coupling as the axion,
$\mathcal{L} = -\frac{1}{4}g_{a\gamma}F_{\mu\nu}\tilde{F}^{\mu\nu}
=g_{a\gamma}\vec{E}\cdot\vec{B}$, are collectively known as axion-like
particles (ALPs). In the case of the QCD axion, the two-photon
vertex is inherited from the two-gluon vertex, thus fixing the relation of the
mass and decay constant as
$|g_{a\gamma}|\mathrm{GeV}\approx 10^{-16}m_a/\mathrm{\mu eV}$
up to a $\mathcal O(1)$ constant~\cite{GrillidiCortona:2015jxo}
(see Ref.~\cite{DiLuzio:2020wdo} for a recent review on axion models).
While ALPs do not solve the strong CP problem, they are theoretically
well motivated as they arise naturally in string theories and
other extensions of the Standard Model~\cite{Arvanitaki:2009fg,Cicoli:2012sz}.

Most axion and ALP searches are based on their two-photon coupling,
see e.g.\ Refs.~\cite{Irastorza:2018dyq, Choi:2020rgn} for recent reviews.
Such a coupling leads to a
conversion between photons and axions in the presence of an
external magnetic field. This phenomenon has been utilised in, e.g.,
the ADMX haloscope~\cite{ADMX:2019uok} and the CAST
helioscope~\cite{CAST:2017uph}
experiments which aim to reconvert respectively DM and solar axions in
the fields of strong magnets. The most extensive limits on the coupling at
sub-eV masses, $g_{a\gamma} < 6.6\times 10^{-11}$, are currently set by
CAST ($m_a\lesssim \unit{eV}$)~\cite{CAST:2017uph} and by studying the
lifetime of stars in the horisontal branch
($m_a\lesssim \unit{keV}$)~\cite{Ayala:2014pea}. The planned ``shine light through a wall''
experiment ALPS-II \cite{Bahre:2013ywa} and solar axion experiment
IAXO \cite{IAXO:2019mpb} are expected to improve these limits significantly.
At present, however, the possible mass of the QCD axion is practically
unconstrained at sub-eV masses. An exception is
the ADMX haloscope which excludes some parts of the QCD axion parameter space
around $\mathrm{few}\unit{\micro eV}$ under the condition that axions account
for the observed dark matter.  The planned
ABRACADABRA~\cite{Ouellet:2018beu} experiment is expected to improve
limits on axionic dark matter immensely, while the IAXO experiments will probe
QCD axions in the 1\,meV$\sim$1\,eV mass range~\cite{IAXO:2019mpb}.

The leading limits on $g_{a\gamma}$ for ALPs with $m_a\lesssim 10^{-6}$\,eV,
$g_{a\gamma}\lesssim (10^{-13}\text{--}10^{-11})$\,GeV$^{-1}$,
are currently set by
astrophysical observations which utilise that the two-photon
coupling leads to a number of interesting changes in gamma-ray spectra.
Most notably, photon-ALP oscillations may effectively increase the
mean-free path of photons in the extragalactic background light (EBL),
since ALPs
travel practically without any interactions~\cite{DeAngelis:2007dqd}.
Additionally, photon-axion oscillations may lead to characteristic
features in the spectra of astrophysical sources of high-energy photons.
The most stringent limits on $g_{a\gamma}$ for
sub-$\mathrm{\micro eV}$ masses  rely on the non-observation of such
spectral signatures and are derived from, e.g.,  gamma-ray observations by
HESS~\cite{HESS:2013udx},  Fermi-LAT~\cite{Fermi-LAT:2016nkz}, 
of SN1987A~\cite{Payez:2014xsa}, and X-ray observations of
Betelgeuse~\cite{Xiao:2020pra}, 
the active galactive nuclei NGC1275~\cite{Reynolds:2019uqt},
the cluster-hosted quasar H1821+643~\cite{Reynes:2021bpe}
and super star clusters~\cite{Dessert:2020lil}.
The former two focus on ``irregularities'' induced by photon-axion
oscillations, while the latter three focus on the production of axions through
the Primakoff effect.
A significant improvement is expected from the
upcoming Cherenkov Telescope Array (CTA)~\cite{CTA:2020hii}. The
interpretation of such results depend, however, significantly on the treatment
of the magnetic fields in and around the source as well as in the extragalactic
space~\cite{Kartavtsev:2016doq,Montanino:2017ara,Libanov:2019fzq,%
Carenza:2021alz}. This applies in particular for the turbulent component of the
magnetic field, for which often only oversimplified models are used.
Moreover, one should note that large scale spectral features can also be
produced by a number of astrophysical effects, including
e.g.\ an electron-beam driven pair cascade~\cite{Wendel:2021mhi}
or even cascades from primary gamma-rays or
nuclei~\cite{Dzhatdoev:2016ftt}.
It is therefore important to identify the features characteristic for 
axion-photon\footnote{We will from now
  on refer to both axions and ALPs simply as `axions'.}  oscillations.

In this work, we study the effect of photon-axion oscillations in a
Monte-Carlo framework based on the {\tt ELMAG}\footnote{The code used in this
  work will be made publicly available in a future release of {\tt ELMAG}.}
code~\cite{Kachelriess:2011bi, Blytt:2019xad},
which is a Monte Carlo program made to simulate electromagnetic cascades
initiated by high-energy photons interacting with the extragalactic background
light.
The use of {\tt ELMAG} allows
us to include properly the interplay of cascading and oscillations, and 
in addition {\tt ELMAG} provides tools to generate turbulent magnetic fields.
We consider for concreteness only a turbulent 
  extragalactic magnetic field, but {\tt ELMAG} can use any magnetic field
  as input.
We discuss the
characteristic signatures expected in the photon spectra from distant
gamma-ray sources, and using domain-like and Gaussian turbulent fields as
examples, show that the predicted signatures
depend rather strongly on the chosen magnetic field model.
As a result, we argue that while the application of domain-like magnetic fields
may be tenable for some quantitative discussions, it should be abandoned in
qualitative studies.
Finally, we propose the use of the discrete power spectrum to detect
photon-axion oscillations in upcoming gamma-ray experiments such as CTA. This
method directly uses the expected characteristic signatures as observable,
namely the energy-dependent wiggles in the photon spectra induced by
photon-axion oscillations. We show that these signatures can in principle be
used to infer information about the magnetic field environment.
While we focus oscillatory features in photon spectra, which we call axion
wiggles, we  comment also on the effect of photon-axion oscillations on the
opacity of the Universe. In particular, we show that the apparent
size of this effect depends strongly on the approximation used for the
turbulent magnetic field.

This paper is structured as follows: In section~\ref{sec:theory}, we present the
numerical treatment of photon-axion oscillations and our Monte Carlo
implementation. Next,  we discuss in section~\ref{sec:magnetic_fields} the
treatment of turbulent magnetic fields. Thereafter, in
section~\ref{sec:params}, we discuss the characteristic oscillatory features
produced by
photon-axion oscillations in the photon spectra and their dependence on the
modelling of the magnetic field, followed by examples in
section~\ref{sec:examples}. In section~\ref{sec:detection},  we present the
suggestion to use the energy dependence of the oscillatory features as
observable in the detection of axions.
Finally, we comment on the importance of a proper treatment of
the magnetic fields when considering the opacity of the Universe, before we
conclude in section~\ref{sec:conclusion}.


\section{Numerical implementation of photon-axion oscillations}
\label{sec:theory}

The physics underlying photon-axion oscillations is discussed clearly in
the classic paper by Raffelt and Stodolsky~\cite{Raffelt:1987im}. Here, we
only highlight the main features of the Lagrangian\footnote{We use
  rationalised natural units with $\hbar=c=1$ and
  $\alpha=e^2/4\pi\simeq 1/137$. Then the critical magnetic field is
  given by  $B_\mathrm{cr}=m_e^2/e\simeq  4.414\times 10^{13}$\,G.}
describing the low-energy interactions between axions, photons and
vacuum fluctuations of electrons,
\begin{equation}
\begin{aligned}
	\mathcal{L} = \mathcal{L}_{aa} &+ \mathcal{L}_{a\gamma} +
	\mathcal{L}_{\gamma\gamma} = \frac{1}{2}\partial^\mu a\partial_\mu a -
	\frac{1}{2} {m_a}^2a^2 -
	\frac{1}{4}g_{a\gamma}F_{\mu\nu}\tilde{F}^{\mu\nu}a \\
	&-\frac{1}{4}F_{\mu\nu}F^{\mu\nu}+
	\frac{\alpha^2}{90{m_e}^4}\left[\left(F_{\mu\nu}F^{\mu\nu}\right)^2 +
	\frac{7}{4}(F_{\mu\nu}\tilde{F}^{\mu\nu})^2\right].
\end{aligned}
\label{eq:lagrangian}
\end{equation}
The first two terms describe the axion $a$ as a free scalar field 
with mass $m_a$, while the third term includes the interaction of
axions with photons, which in the presence of an external magnetic
field will result in axion-photon oscillations. The last term is the
Euler-Heisenberg effective Lagrangian that takes into account 
vacuum polarisation effects below the creation threshold of
electron-positron pairs. In particular, this term
leads to a refractive
index for photons in an external electromagnetic field which influences the
propagation and oscillation of axions and photons. Following the
usual convention of the wave vector $\vec{k}$ pointing in the direction of
the photon electric field, the refractive indices of the photon in the
longitudinal ($\parallel$) and transverse ($\perp$) directions are given by
\begin{equation}
	n_\perp = 1 + \frac{4}{2}\xi \qquad \text{ and } \qquad
  n_\parallel = 1 + \frac{7}{2}\xi
  \label{eq:photon_refractive_index}
\end{equation}
with $\xi\equiv (\alpha/45\pi)(B_\perp/B_\mathrm{cr})^2$.

The Lagrangian \eqref{eq:lagrangian} leads after linearisation
to the equation of motion
\begin{equation}
	\left(E + \mathcal{M} - \iu \partial_z\right) \phi(z) = 0,
	\label{eq:eom}
\end{equation}
where we denote the energy of photons and axions by $E$,
have chosen the $z$-axis as propagation direction  
and have introduced the wave function
\begin{equation}
	\phi(z) = \mqty(A_\perp\\A_\parallel\\a).
\end{equation}
The mixing matrix is given by
\begin{equation}
	\mathcal{M}=
 	\mqty(\Delta_\perp & 0 & 0 \\
	 	0 & \Delta_\parallel & \Delta_{a\parallel} \\ 	
	 	0 & \Delta_{a\parallel} & \Delta_{a}),
  \label{eq:mixing_matrix}
\end{equation}
where $\Delta_{\|,\perp}=(n_{\|,\perp}-1)E$ and
$\Delta_{a} = -{m_a}^2/(2E)$.
The two polarisation states of the photon are given as
linear polarisation states perpendicular
and parallel to the transverse magnetic field at a given position.
The off-diagonal terms lead to photon-axion mixing in the presence of an
external magnetic field,
\begin{equation}
	\Delta_{a\parallel} = \frac{g_{a\gamma}}{2} B_\perp.
  \label{eq:mixing_term}
\end{equation}
In general, the diagonal terms $\Delta_{\|,\perp}$ in Eq.~\eqref{eq:eom} should
include the total refractive index of the photon. Other contributions
describing the photon dispersive effects of the medium and the EBL,
as well as the chosen numerical values,
will be discussed in section~\ref{sec:params}.

It is useful to consider the propagation through a homogeneous magnetic
field to obtain an understanding of the problem.
In this case, Eq.~\eqref{eq:eom}  simplifies to
\begin{equation}
	\left[E - \iu \partial_z +
	\mqty(\Delta_\| & \Delta_{a\|} \\ \Delta_{a\|} & \Delta_{a})\right]
	\mqty(A_\| \\ a) = 0.
\end{equation}
The photon conversion probability then becomes
\begin{equation}
	P_s(\gamma\to a) = \left|\braket{A_\|(0)}{a(s)}\right|^2=
  \left(\Delta_{a\|}s\right)^2
	\frac{\sin^2(\Delta_\mathrm{osc}s/2)}{(\Delta_\mathrm{osc}s/2)^2}
  \label{eq:homogeneous_solution}
\end{equation}
with
\begin{equation}
	\Delta_\mathrm{osc}^2 = (\Delta_\| - \Delta_a)^2 + 4\Delta_{a\|}^2.
\end{equation}
Similarly, the oscillation length in any sufficiently smooth
environment is given by $L_\mathrm{osc}\simeq 2\pi/\Delta_\mathrm{osc}$.
Thus, one can see that the oscillation length, the correlation length
$L_\mathrm{c}$ of the magnetic field and the mixing strength
$2\pi/\Delta_{a\|}$ are the main parameters determining the effects of
photon-axion oscillations. For example, when
$\Delta_\mathrm{osc}\sim 2\Delta_{a\|}$, we enter the strong mixing regime where
Eq.~(\ref{eq:homogeneous_solution}) gives
$P_s(\gamma\to a)=\sin^2(\Delta_{a\|} s)$.

We describe now how the photon-axion equation of motion \eqref{eq:eom}
is implemented into  {\tt ELMAG}~\cite{Kachelriess:2011bi,Blytt:2019xad}.
For convenience, we will refer to the superposition of a photon and an axion
as a `phaxion'. The probability
that the phaxion interacts with the EBL at position $s$ is
$\dd P=P_s(\gamma\to\gamma)\sigma_{\rm pair}(s)\dd{s}$ with
$P_s(\gamma\to\gamma)=1-P_s(\gamma\to a)$. This is equivalent to
checking
first whether a photon would interact at that position and next to account
for the probability that a phaxion would be measured as a photon. 
This motivates the following numerical scheme which takes into account
the absorption of photons:
\begin{enumerate}
  \item Start with a pure photon with an energy drawn from
    the distribution for the injection energy.
    For an unpolarised gamma-ray source, choose randomly a linear
    polarisation state.
  \item Draw the interaction length of a phaxion, $\lambda$, at the current
    position in accordance to the mean free path length a photon.
  \item Propagate the phaxion from $s$ to $s+\lambda$ according to the
    phaxion equation of motion.
  \item If $P_{\gamma\gamma} > r$ for a random number $r$ chosen from a uniform
    distribution $r\in[0, 1]$, the phaxion wave function collapses into a photon
    and the photon undergoes pair production in interaction with the EBL.
    If not, go to point 2.
  \item Treat the electromagnetic cascade that arises, and for each photon
    go to point 2.
\end{enumerate}

The Monte Carlo treatment of the photon-axion oscillations implemented
in this work has several advantages compared to conventional (semi-)analytical
approaches (see e.g.\ Ref.~\cite{Galanti:2018nvl} and references therein) and
the procedure in Ref.~\cite{Meyer:2021pbp}, at the cost of being
computationally more demanding. First, the implementation of realistic magnetic
fields and additional effects like an inhomogeneous electron density is
trivial. Second, photon absorption  can be considered on an event-by-event
basis and the resulting electromagnetic cascade can be accounted for. Third,
polarisation effects are by default included. Finally, this method allows to
include the effect of photon-axion oscillations into other studies of
electromagnetic cascades. For example,  a potential increase in the size of
gamma-ray halos  around astrophysical sources because of
the increased mean-free path  of photon could be studied using {\tt ELMAG}
in a straight-forward way, see e.g.\ Ref.~\cite{Batista:2021rgm} for a
recent review on the subject.


\section{Turbulent magnetic fields}
\label{sec:magnetic_fields}

We describe in this work turbulent magnetic fields as   isotropic,
divergence-free Gaussian random fields with zero mean, rms value
$B_\mathrm{rms}$ and zero helicity. The algorithm implemented in
{\tt ELMAG} for the generation of such fields is based on the method
suggested in Refs.~\cite{1994ApJ...430L.137G,1999ApJ...520..204G}
and described in Ref.~\cite{Blytt:2019xad}. In this approach,
the turbulent magnetic field is modelled as a superposition of $n$ left-
and right-circular polarised Fourier modes. The spectrum of the modes
is chosen as a power-law
\begin{equation}
  B_j = B_\mathrm{min}(k_j/k_\mathrm{min})^{-\gamma/2}
\end{equation}
between $k_\mathrm{min}$ and $k_\mathrm{max}$, 
corresponding to the largest $L_\mathrm{max}=2\pi/k_\mathrm{min}$
and smallest scales $L_\mathrm{min}=2\pi/k_\mathrm{max}$, respectively.
The quantity  $B_\mathrm{min}$ is fixed by normalising the total field
strength to $B_\mathrm{rms}$.
The coherence length is in turn connected to $L_\mathrm{max}$ and
$L_\mathrm{min}$ as 
\begin{equation}
  L_\mathrm{c} = \frac{L_\mathrm{max}}{2}\frac{\gamma - 1}{\gamma}
  \frac{1-(L_\mathrm{min}/L_\mathrm{max})^\gamma}
  {1-(L_\mathrm{min}/L_\mathrm{max})^{\gamma-1}}
  \simeq \frac{L_\mathrm{max}}{2}\frac{\gamma-1}{\gamma},
\end{equation}
the last equality being valid for $L_\mathrm{min}\ll L_\mathrm{max}$.
For definiteness, we will consider a time-independent $B_\mathrm{rms}$
  and a Kolmogorov spectrum with $\gamma=5/3$
for which $L_\mathrm{c}\simeq L_\mathrm{max}/5$.

In the literature,  a so-called domain-like magnetic field has often been
used to describe photon-axion oscillations in turbulent astrophysical magnetic
fields,
see e.g.~\cite{Grossman:2002by,Wouters:2012qd,DeAngelis:2011id,%
Meyer:2014epa,Galanti:2018nvl,Galanti:2018myb,Galanti:2018upl,%
Reynolds:2019uqt,Buehler:2020qsn,Reynes:2021bpe,Meyer:2021pbp}.
In this approach, the magnetic field along the line of sight is
divided into patches with size equal to the coherence length
$L_\mathrm{c}$ of the field. The field in each patch is assumed to
be homogeneous, while its direction is chosen randomly. Such an
approximation certainly breaks down  when the oscillation
length becomes
smaller than the coherence length, $L_\mathrm{osc}\lsim L_\mathrm{c}$:
In this case, oscillations probe the magnetic field structure on scales smaller
than the domain size, which are neglected in this simple model. Thus, a more
realistic description of the turbulent magnetic field, including fluctuations
on various scales, should be used.

\begin{figure}
  \centering
  \includegraphics[scale=1]{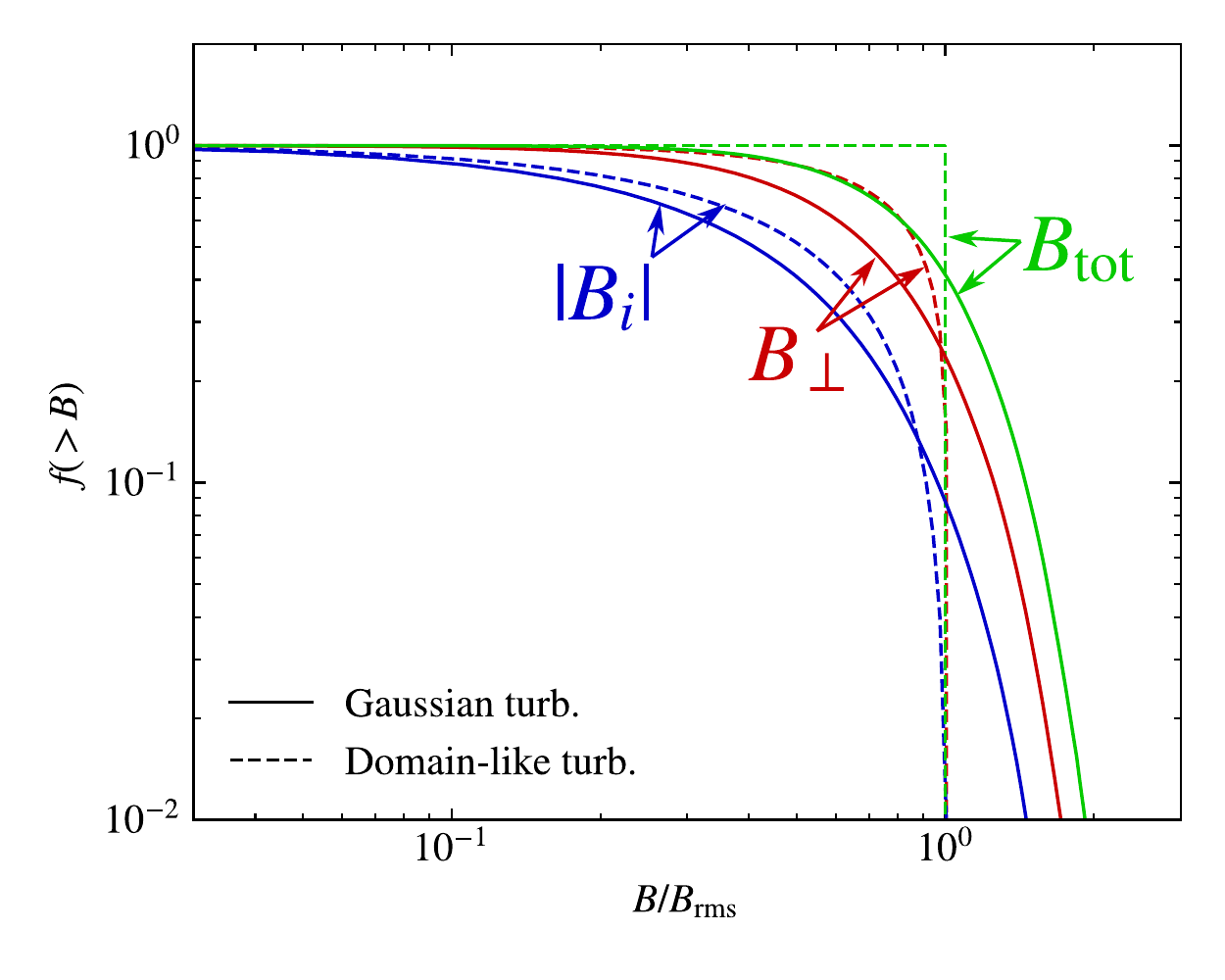}
  \caption{The cumulative distribution $f(B)$ of the magnetic field strengths
    for the turbulent (solid lines) and domain-like (dashed lines)
    cases are compared.
    The total magnetic field strength is shown in green, transverse strength in
    red and the strength in a given direction in blue.
    }
  \label{fig:B_distribution}
\end{figure}

In the case of a large oscillation length, $L_\mathrm{osc}\gsim L_\mathrm{c}$,
on the other hand, one may expect the approximation of domains to be valid
since the power of the turbulent magnetic field is contained mainly in its
large-scale modes for $\gamma> 1$. However, also in this limit the
large-scale fluctuations of the turbulent field  will lead to important
differences in the photon conversion relative to the domain-like case.
To visualise these effects, the cumulative distribution of the magnetic
field, i.e.\ the fraction $f(>B)$ of volume filled with a magnetic field
stronger than $B$, is compared for the two models
in Fig.~\ref{fig:B_distribution}.%
\footnote{Since the distribution of the magnetic fields for the Gaussian
  turbulent field and domain-like turbulence only depends on $B_\mathrm{rms}$,
  Fig.~\ref{fig:B_distribution} remains unchanged if the coherence length
  $L_c$ or the slope $\gamma$ is changed.
}
Since the fluctuations of the turbulent magnetic field are
spread over a large range of scales, the variance of the perpendicular
field is greater and extends to larger field strengths
than in the domain-like case. Considering single field realisations---as
should be done for the study of single sources---in the domain-like
case thus underestimates the expected ``cosmic variance''.

\begin{figure}
  \centering
  \includegraphics[scale=1]{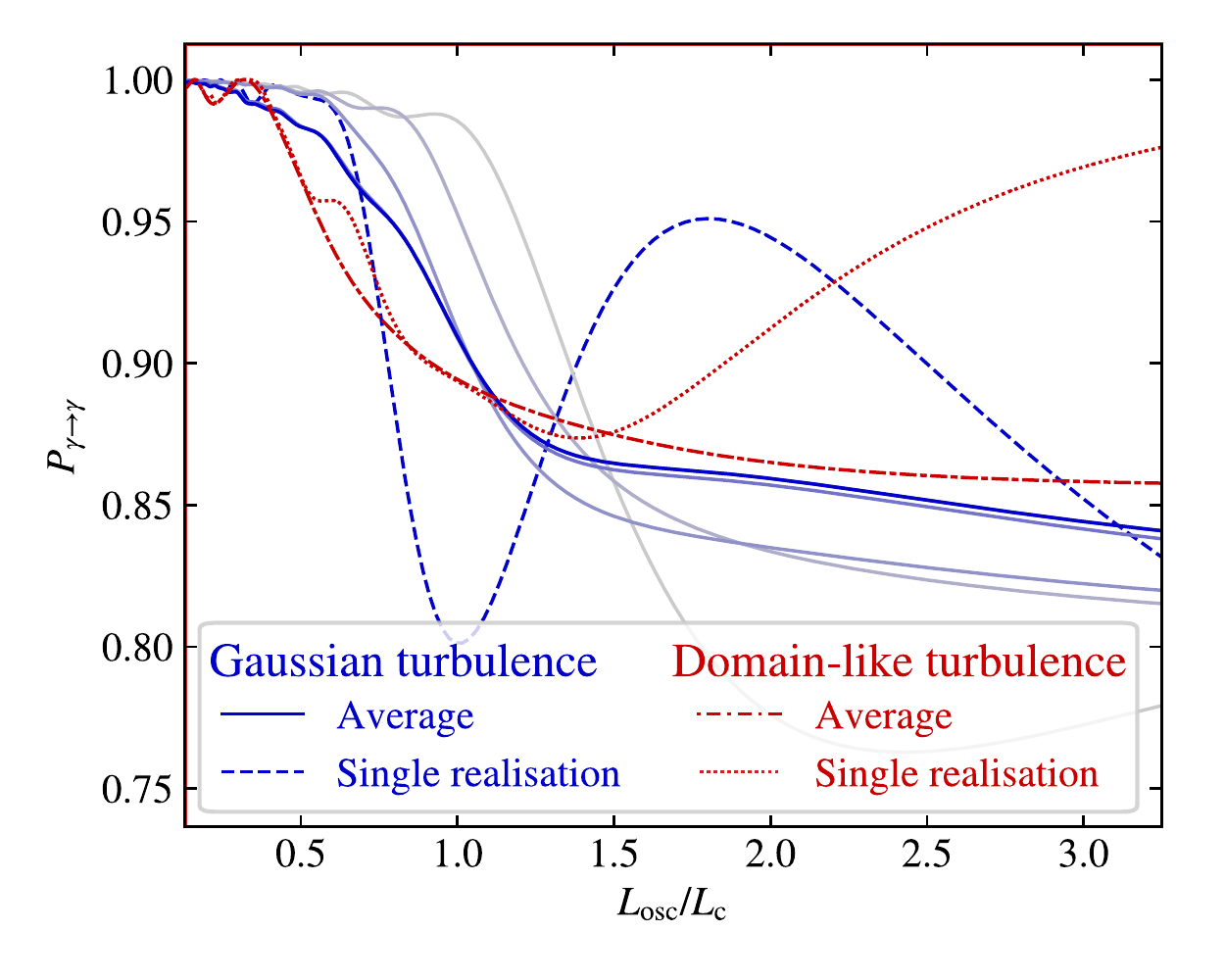}
  \caption{The average photon survival probability after 12 coherence lengths
  as a function of the
  oscillation length using 100 realisations is shown in for
  Gaussian (blue solid lines) and
  domain-like (red dashed dotted lines) turbulence.
  The oscillation length
  is computed by setting $B_\perp\simeq B$, and the parameters
  $m_a=10^{-10}\unit{eV}$, $g_{a\gamma}=10^{-16}\unit{eV^{-1}}$,
  $B_\mathrm{rms}=B_\mathrm{tot}=10^{-12}\unit{G}$ and
  $L_\mathrm{c}=2\unit{Mpc}$ are used. The
  results for a single realisation are shown in dashed lines
  and dotted lines.
  The opaqueness of
  the blue  lines indicate the number of modes $n_k$ included in the generation
  of the Gaussian turbulence as described in the text;
  $n_k=\{2, 10, 20, 50, 100\}$ from lightest to darkest.
  }
  \label{fig:nmode}
\end{figure}

The effects discussed above
are clearly visible in Fig.~\ref{fig:nmode}, where we plot the photon survival
probability as a function of the oscillation length for a propagation
distance equal to twelve coherence lengths choosing the
parameters\footnote{We set
$L_\mathrm{min}=0.01\unit{Mpc}$ and $L_\mathrm{max}=5\unit{Mpc}$.} of
the extragalactic field as
$B_\mathrm{rms} = B_\mathrm{tot} = 10^{-12}\unit{G}$ and
$L_\mathrm{c}=2\unit{Mpc}$, while the axion parameters were set to
$m_a=10^{-10}\unit{eV}$, $g_{a\gamma}=10^{-16}\unit{eV^{-1}}$.
The average photon survival probability from 100~realisations for a 
turbulent (blue solid) and a domain-like field (red solid) are shown. The
opaqueness of the blue lines indicates the number of modes included;
$n_k=\{2, 10, 20, 50, 100\}$ from the lightest to the darkest line.
The result for a single
realisation of the magnetic field is shown in dashed lines for comparison.
The general energy dependence of the survival probability is similar for
the two magnetic field models: Below a given threshold energy determined by
$\Delta_\mathrm{osc}\sim \Delta_{a\gamma}$, the survival probability is
close to unity. Above the threshold energy, we are in the strong mixing
regime. There are, however, clear differences in the detailed behaviour.
Most notably, the oscillations for a turbulent field are smoothed out
by the variation in the transverse magnetic field strength compared to
the domain-like case. Furthermore, the  survival probabilities predicted
using the two magnetic field models differ even in
the case when $L_\mathrm{c}>L_\mathrm{osc}$, as expected from the discussion
in the previous paragraph. Meanwhile, the change induced by the
increase  of the number of modes\footnote{The
  normalisation is kept constant while additional modes are added towards 
  smaller scales. Therefore, the total energy stored in the magnetic
  field increases slightly with increasing number of included modes in
  this example.}   shows the importance 
of an accurate description of magnetic field also on small-scales.
It is interesting to note that Ref.~\cite{Galanti:2018nvl} introduced a linear
interpolation between the magnetic fields in the domain-like approach in order
to resolve the discontinuities in the domain-like magnetic fields. While this
approach leads to a slight smoothening of the peaks observed in
Fig.~\ref{fig:nmode}
for the domain-like magnetic field, other effects like the variation for
different realisations and the shift in the threshold energy are not
reproduced in this approach.

\section{Parameter space of photon-axion oscillations}
\label{sec:params}

Photon-axion oscillations are essentially determined by the 
axion parameters ($m_a$ and $g_{a\gamma\gamma}$) and the refractive indices
induced by the magnetic field, the medium and the EBL. In addition,
the propagation distance and the photon energy enter the problem. The
effect of the magnetic field via the QED vacuum polarisation,
$\Delta^\mathrm{QED}_{\|,\perp}$, was already discussed in
section~\ref{sec:theory}. Among the medium effects,
we neglect the Faraday contribution as the random direction of the turbulent
magnetic field averages out its effect, as well as the Cotton-Mouton effect.
Then the effective mass of the photon in a plasma,
\begin{equation}
 m_\mathrm{pl} \simeq \omega_\mathrm{pl} = \sqrt{\frac{4\pi\alpha n_e}{m_e}}
  \simeq 0.0371 \factor{n_e}{1}{cm^{-3}}^{1/2} \unit{neV},
\end{equation}
leads to the only additional contribution induced by the medium,
$\Delta_{\|,\perp}^{\rm pl}=-m_{\rm pl}^2/(2E)$ \cite{Raffelt:1987im}. In
addition,
the EBL and starlight may have profound effects on the refractive index
at large energies, as first realised in Ref.~\cite{Dobrynina:2014qba}.
The isotropic EBL influences the two polarisation states equally, and its
contribution is given
by
\begin{equation}
\Delta_\mathrm{EBL}\simeq \Delta_\mathrm{CMB} \simeq 0.522\cdot 10^{-42}E.
\end{equation}
This approximation is valid below the pair creation threshold
$E_{\rm th, CMB}\simeq 400\unit{TeV}$ on CMB photons, which are dominating
the contribution of the EBL to the photon refractive index. In summary, we
take into account the most important additional effects on the
photon refractive index by using $\Delta_{\|,\perp}=
\Delta^\mathrm{QED}_{\|,\perp}+\Delta_\mathrm{CMB}+\Delta_\mathrm{pl}$ in 
the mixing matrix~\eqref{eq:mixing_matrix}.

For the ease of comparison and identification of scales in different
astrophysical environments, we will consider in the following as
numerical values for these free parameters
\begin{equation}
  \begin{aligned}
    \Delta_\parallel^\mathrm{QED} &= 1.5\times 10^{-9}\unit{Mpc^{-1}}
      \factor{E}{10^{11}}{eV}\factor{B_\perp}{10^{-9}}{G}^2, \\
    \Delta_\parallel^\mathrm{pl} &= -1.1\times 10^{-10}\unit{Mpc^{-1}}
      \factor{n_e}{10^{-7}}{cm^{-3}}\factor{E}{10^{11}}{eV}^{-1},\\
    \Delta_\parallel^\mathrm{CMB} &= 8.2\times 10^{-3}\unit{Mpc^{-1}}
      \factor{E}{10^{11}}{eV},\\
    \Delta_a &= -7.8\times 10^{-3}\unit{Mpc^{-1}}\factor{m_a}{10^{-10}}{eV}^2
      \factor{E}{10^{11}}{eV}^{-1},\\
    \Delta_{a\parallel} &= 1.5\times 10^{-2}\unit{Mpc^{-1}}
      \factor{B_\perp}{10^{-9}}{G}\factor{g_{a\gamma}}{10^{-20}}{eV^{-1}}.\\
  \end{aligned}
  \label{eq:params}
\end{equation}
We note that the value of the extragalactic magnetic field---which is
often used in the literature---is close to the limits derived
in Refs.~\cite{Pshirkov:2015tua,Jedamzik:2018itu}.

In general, photon-axion oscillations depend both on the magnetic field
strength and the plasma density. However, we can eliminate one of the
two variables using the conservation of magnetic flux in a plasma. Then
the magnetic field lines are frozen to the fluid elements and, neglecting
dissipation and dynamo effects, we can employ the simple scaling relation%
\begin{equation}
  n_e\simeq n_{e,0}\left(\frac{B}{B_0}\right)^{\eta},
  \label{eq:n_e}
\end{equation}
with $\eta=3/2$ for isotropic volume changes.
For concreteness, we set as reference $B_0=\unit{\micro G}$ and
$n_{e,0}=0.02\unit{cm^{-3}}$ which is suitable for the Milky
Way\footnote{Although we focus on extragalactic environments in this work, the
galactic environment is put as reference since plasma effects are negligible
for extragalactic propagation, as we soon will see.
However, for 
magnetic fields $B\sim (10^{-9}\text{--}10^{-10})\unit{G}$ one
obtains $n_e\sim (6\times 10^{-7}\text{--}2\times 10^{-8})\unit{cm^{-3}}$, 
suitable for extragalactic space.
}~\cite{Ferriere:2001rg}.
Although this scaling should not be considered a general
rule, 
it is sufficient for the purposes in this paper.

From the homogeneous solution~\eqref{eq:homogeneous_solution}, one can further
conclude that the photon conversion probability
will be governed by the relative ratios of $\Delta_\mathrm{osc}^{-1}$,
$\Delta_{a\|}^{-1}$, $L_\mathrm{c}$ and the distance  $s$ travelled.
That is, in order to have a significant conversion of photons, one must have
a sizeable amount of oscillations ($s\Delta_\mathrm{osc}\gtrsim 1$) and a 
sizeable mixing ($\Delta_{a\|}\sim \Delta_\mathrm{osc}$).
The coherence length, meanwhile, determines the intrinsic behaviour of the
conversion probability: If $L_\mathrm{c}\gg 2\pi/\Delta_\mathrm{osc}$ the
conversion probability ``probes'' the magnetic field with several oscillations
per coherence length and the photon state parallel to the transverse magnetic
field is completely mixed with the axion for each coherence length. If
$L_\mathrm{c}\ll 2\pi/\Delta_\mathrm{osc}$, on the other hand, the magnetic
field changes quickly so that the mixing slowly converges.
Observationally, one can measure the energy spectrum of single gamma-ray
sources, which means that one can probe the energy dependence of the
photon-axion oscillation probability. The only energy dependence of the
characteristic parameters lies in $\Delta_\mathrm{osc}$ and its generic
behaviour is the
same for all astrophysical environments (see also Ref.~\cite{Galanti:2018nvl}
for a similar discussion):
\begin{enumerate}
  \item $\Delta_\mathrm{osc}\sim E^{-1}$ at low energies. Here, the oscillation
    length is determined by the effective photon mass or the axion mass,
    depending on the magnetic field strength and the axion
    parameters.
  \item $\Delta_\mathrm{osc}\sim E^0$ at intermediate energies. This is the 
    strong mixing regime where the oscillation length is determined by the
    mixing term, $\Delta_{\|}\sim 2\Delta_{a\|}$.
  \item $\Delta_\mathrm{osc}\sim E^1$ at large energies. The oscillation length
    is here determined by either the CMB or the vacuum polarisation depending
    on the magnetic field strength. 
\end{enumerate}
The transitions between these regimes occur around the energies
$E_\mathrm{min}$ and $E_\mathrm{max}$ defined by
$4\Delta_{a\|}=(\Delta_\|-\Delta_a)^2$. Depending on the treatment of
the magnetic fields, the oscillation probability in the transition region
vary. For instance, the larger variance in $B_\perp$ seen in
Fig.~\ref{fig:B_distribution} for a turbulent field
will lead to a larger variance in the threshold energies $E_\mathrm{min}$ and
$E_\mathrm{max}$. This will effectively reduce or even cancel oscillations
close to the thresholds upon averaging and shift the threshold 
energies, as already seen in Fig.~\ref{fig:nmode}. In
section~\ref{sec:detection}, we will discuss how these generic energy
dependence can
be used as an observable in the search for axion-photon oscillations in
astrophysical environments.

\begin{figure}
\centering
\includegraphics[scale=1]{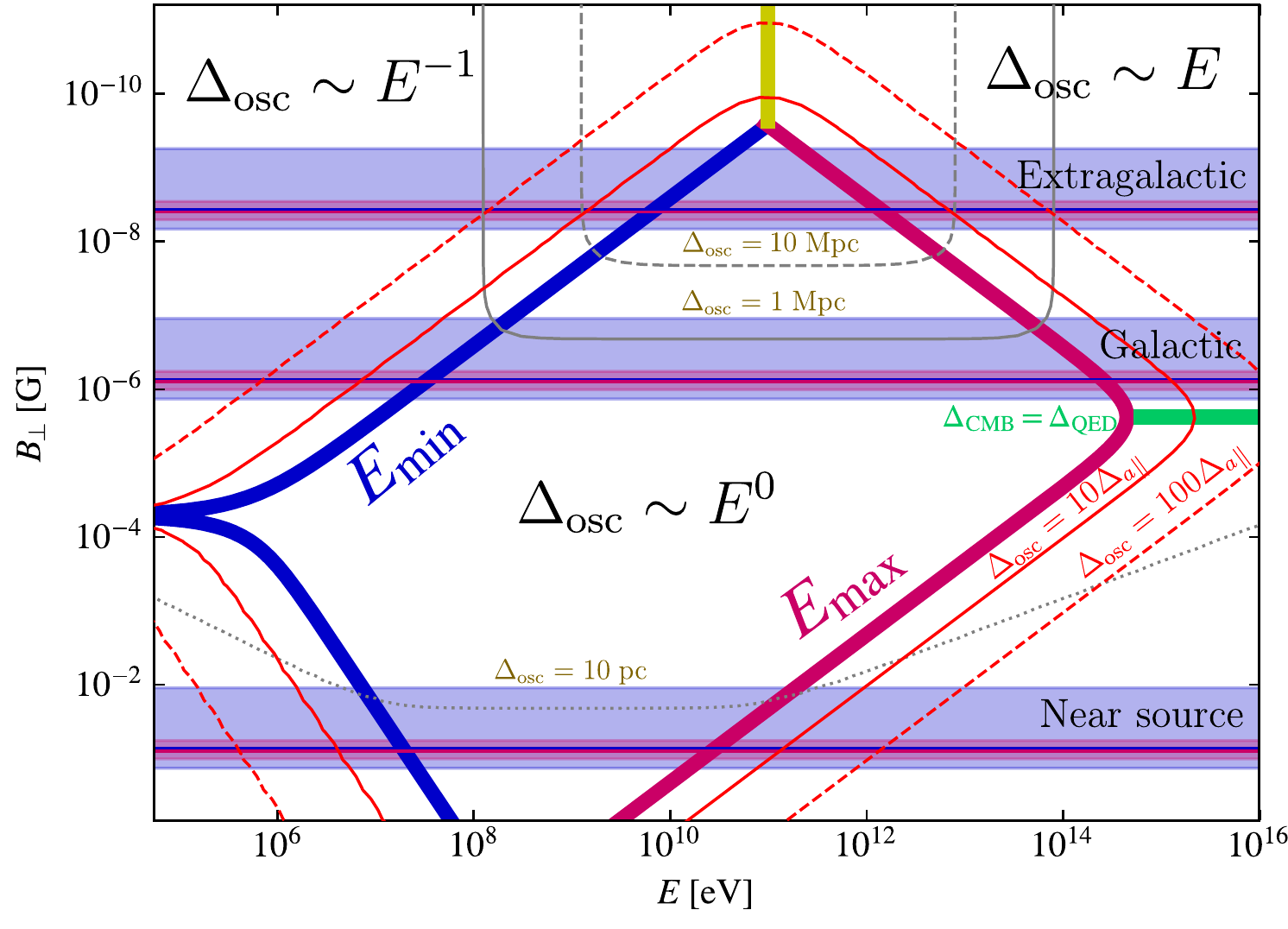}
\caption{
Visualisation of the photon-axion parameter space as explained in the text.
The space is split into five main parts depending on the main contribution
to the oscillation wave number $\Delta_\mathrm{osc}$: $\Delta_a$ (upper left),
$\Delta_\mathrm{pl}$ (lower left), $\Delta_{a\|}$ (middle),
$\Delta_\mathrm{CMB}$ (upper right) and $\Delta_\mathrm{QED}$ (lower right).
The transition to the strong mixing regime
$\Delta_\mathrm{osc}\sim\Delta_\mathrm{a\|}$ occurs at $E_\mathrm{min}$
and $E_\mathrm{max}$ as indicated in blue and red thick lines, respectively.
The value of $\Delta_\mathrm{osc}$ is shown in brown lines for $10\unit{pc}$,
$1\unit{Mpc}$ and $10\unit{Mpc}$, while its value in terms of
$\Delta_{a\|}$ is shown in red lines.
Furthermore, the magnetic field strength and variation for Gaussian and
domain-like turbulences for extragalactic, galactic and ``source near''
environments are shown in shaded blue and red regions. For convenience, we have
used $B\simeq B_\perp$ in this plot.
}
\label{fig:param2}
\end{figure}

The discussions above can be summarised in the energy-magnetic field plane, as
shown in Fig.~\ref{fig:param2} choosing the axion parameters\footnote{Note
  that a change in the magnetic field strength is equivalent to
  the inverse change of the coupling strength
  (with the exception of the contribution from the plasma),
  since the product of these
  two quantities determines the oscillation amplitude.
} as in
Eq.~\eqref{eq:params}.
This plot is essentially a representation of
$\Delta_\mathrm{osc}$ as function of energy and the transverse magnetic field
strength. The red lines indicate the value of
$\Delta_\mathrm{osc}$ in terms of $\Delta_{a\|}$, and the brown lines its value.
The transverse magnetic field value and its $1\sigma$ distribution (from
Fig.~\ref{fig:B_distribution}) is shown for three field strengths
indicating the upper limit for extragalactic fields, a typical value for
the Galactic magnetic field  and for field close to sources.
The three regions, $\Delta_\mathrm{osc}\sim E^{-1}$,
$\Delta_\mathrm{osc}\sim E^{0}$ and $\Delta_\mathrm{osc}\sim E^{1}$, are
indicated  in the figure. Furthermore, the parameter space
is divided into five regions depending on the dominant contributions to
$\Delta_\mathrm{osc}$: axion mass (upper left), plasma (lower left), mixing
(middle), CMB (upper right) and QED (lower right). For very weak magnetic field
strengths, there is no strong mixing regime and so the transition is marked
in yellow. Likewise, the transition from a CMB to a QED dominated refractive
index is shown in green. As an example of how this plot can be used, consider
the extragalactic magnetic field: At low energies, the oscillations will be
governed by the axion mass term $\Delta_a$. In the energy range
$E=10^9\text{--}10^{12}\unit{eV}$, we are in the strong mixing regime where
there are no energy-dependent oscillations. The exact energy of this
transition will vary by
around an order of magnitude for different realisations of the Gaussian
turbulence.
The energy oscillations will occur close to, but outside, the strong mixing
regime, and their strength depends on $\Delta_\mathrm{osc}/\Delta_{a\|}$.
These features will be shown explicitly with an example in the next section.
A short discussion on how Fig.~\ref{fig:param2} changes with $m_a$ is
  given in the appendix.

An interesting region in parameter space is  where the plasma contribution
cancels (on average)
the axion contribution, $|\Delta_a| = |\Delta_\mathrm{pl}|$. According to the
previous discussions, the strong mixing regime will then extend to arbitrary
low energies, potentially reaching the CMB. The homogeneity of the CMB
can in principle thus be used to set stringent bounds on $g_{a\gamma}$ for
specific $m_a$. Interestingly, from Fig.~\ref{fig:param2} one
notes that this ``resonance'' transition will always occur in the passage
from a typical gamma-ray source to its surrounding with weaker magnetic fields.
It should be noted, however, that the resonance region should 
be comparable to the oscillation length in order to have detectable effects.
Such a resonance transition in the early Universe has previously been 
discussed in Ref.~\cite{Mirizzi:2009nq}, where the homogeneity of the CMB was
used to set limits on the axion parameters.

\section{Characteristic axion signatures}
\label{sec:examples}

In order to apply the concepts discussed in the previous sections, we consider
now two concrete, albeit over-simplified, examples. A more realistic example
will be considered in the next section, wherein we introduce a method of
detecting the characteristic wiggles in the photon spectrum from a physical
source. 

\begin{figure}
  \centering
  \includegraphics[scale=1]{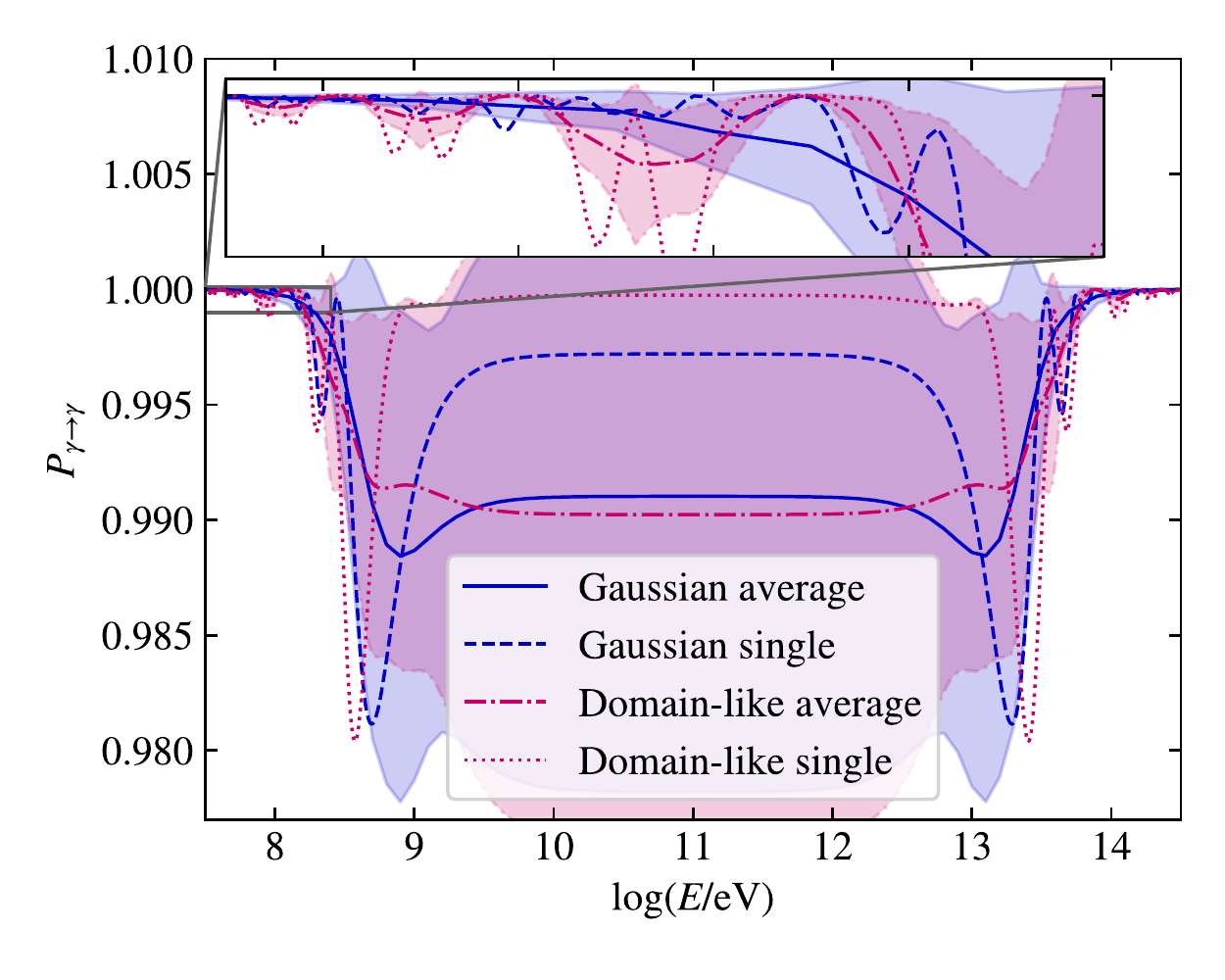}
  \caption{The photon survival probability for polarised photons propagating
  through five coherence lengths is plotted as a function of energy.
  The coherence length is set to $L_\mathrm{c}=1\unit{Mpc}$, and the axion
  parameters and extragalactic magnetic field parameters in
  Eq.~\eqref{eq:params} are used.
  The average of 100 realisations of the magnetic field is plotted for Gaussian 
  (blue solid) and domain-like turbulence (red dashed dotted), with their
  corresponding standard deviations indicated by the shaded regions. The
  results from a single realisation is shown in dashed and dotted lines,
  respectively. To better visualise the oscillations outside the strong mixing
  regime, a portion of the plot is enlarged.
  }
  \label{fig:energy_dependence}
\end{figure}

We consider a photon source at a distance of $5\unit{Mpc}$ for the parameters
given in Eq.~\eqref{eq:params}. The resulting photon survival probabilities
using different models for the magnetic field are shown in
Fig.~\ref{fig:energy_dependence} as a function of the photon energy.
Averaging over 100 realisations, we see
the same characteristics as before:
at low and high energies, the photon survival probability is close to unity,
while there is a strong mixing regime at intermediate energies. However, there
are even after averaging  clear differences in the
results caused by the treatment of the magnetic field. Importantly, the
variation in the survival probability between realisations is much larger
using a turbulent field than in the domain-like treatment, in accordance
to the variation in the magnetic field itself shown in
Fig.~\ref{fig:B_distribution}. Looking at single realisations, one can
clearly see the effect of the energy-dependent oscillation length. That is,
the energy spectrum will have wiggles with a wave number scaling with
$\Delta_\mathrm{osc}$, where the energy dependence of $\Delta_\mathrm{osc}$
at low energies is $\Delta_\mathrm{osc}\sim E^{-1}$, then
$\Delta_\mathrm{osc}\sim\mathrm{const.}$
and finally at high-energies $\Delta_\mathrm{osc}\sim E^{1}$.
We note in particular that even though the oscillations on average cancel in
a turbulent magnetic field, the oscillations in a single realisation---which
is the relevant case for observations---may be huge.

\begin{figure}
  \centering
  \includegraphics[scale=1]{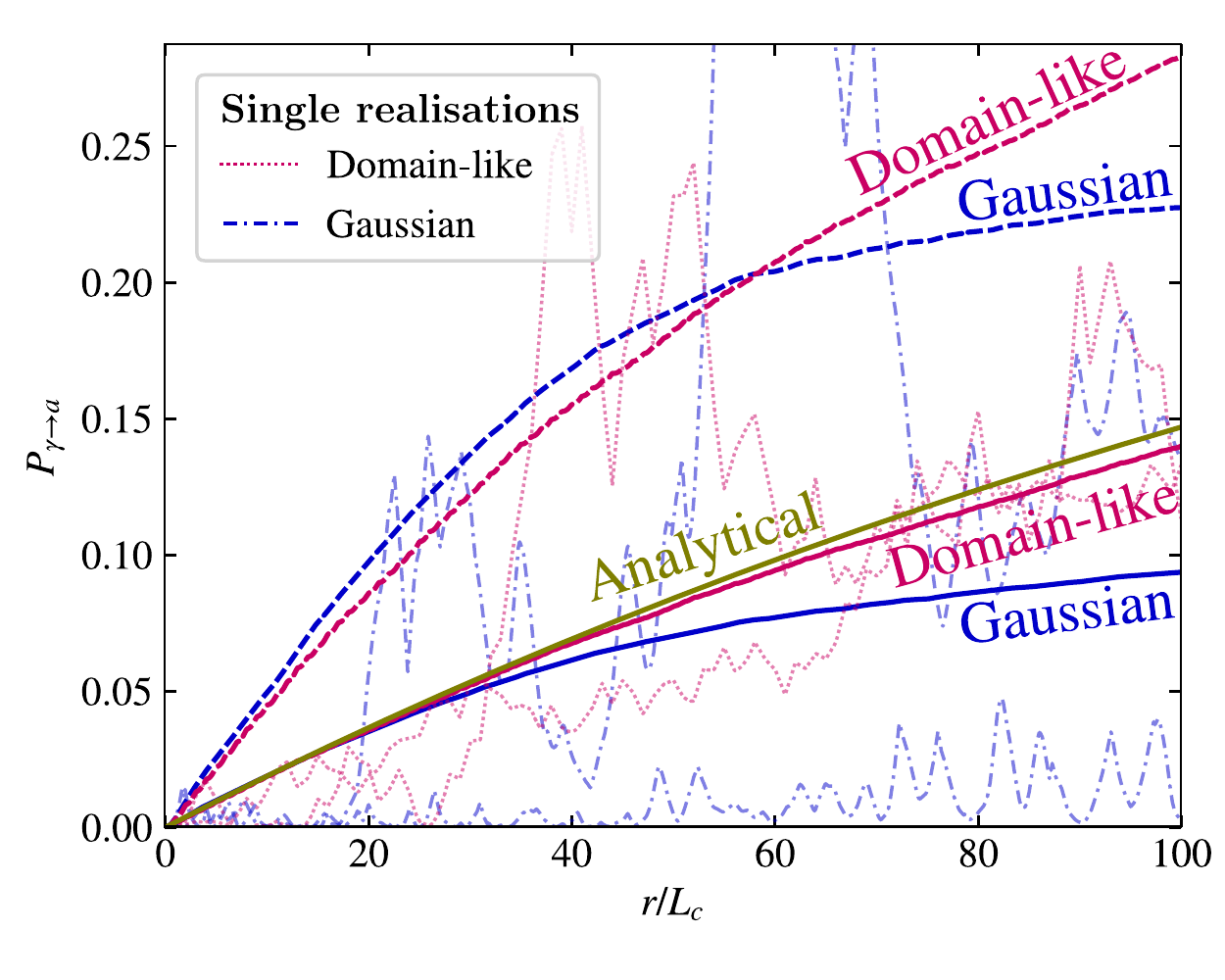}
  \caption{
  The average photon survival probability for polarised photons
  is plotted as a function of distance for $10^4$ realisations of the magnetic
  fields and a fixed energy, $E=10^{11}\unit{eV}$.
  The same parameters and color scheme as in
  Fig.~\ref{fig:energy_dependence} are used. The solid lines indicate the
  averaged results, while the dashed lines indicate their $1\sigma$ variation.
  Furthermore, the results from two single realisations are shown in
  dashed dotted (Gaussian) and dotted (domain-like)
  lines for visualisation. Finally, the analytical approximation
  (Eq.~\eqref{eq:domain-like-approx}) is shown in yellow for comparison.
  }
  \label{fig:distance_dependence}
\end{figure}

Next, we consider in Fig.~\ref{fig:distance_dependence} the same set-up but
for a fixed energy, $E=10^{11}\unit{eV}$, as function of the distance for $10^4$
realisations of the magnetic field. The average photon conversion
probability increases, as expected, slowly
towards 1/3. Again, the relative variation for different realisations of
the turbulent field is much larger than for the domain-like case, although
the average values are similar. Interestingly, the survival probability
for a single magnetic field realisation can vary almost over all the allowed
range of values.  This implies that the oscillation probability for a
specific source can deviate strongly from the average, as already discussed in
Refs.~\cite{Mirizzi:2009aj,Kartavtsev:2016doq}. Moreover,
the large ``cosmic variance'' prevents to define a characteristic signal
which can be probed as function of the source distance.

An interesting observation is that the convergence time for the turbulent
field is much larger than for the  domain-like one at large distances. At
short distances, the conversion probability is slightly larger for the
turbulent field, cf.\ also Fig.~\ref{fig:energy_dependence}.
This can be understood by the following argument. The average
conversion probability for a photon in a domain-like turbulence with
\emph{constant transverse magnetic field}, $B_\perp$,
is~\cite{Grossman:2002by,Mirizzi:2006zy}
\begin{equation}
  P_{\gamma\to a} = \frac{1}{3}\left(1 - \me^{-r/L_\tau}\right),
  \label{eq:domain-like-approx}
\end{equation}
where $L_\mathrm{\tau}=2L_\mathrm{c}/3P$ and $P$ is given by
Eq.~\eqref{eq:homogeneous_solution}. In Fig.~\ref{fig:distance_dependence},
the strong mixing regime was considered, in which case $P$ depends on the
magnetic field as $P= \sin^2(B_\perp g_{a\gamma}L_\mathrm{c})/4\sim
B_\perp^2 + \mathcal{O}(B_\perp^4)$. The decay rate will thus be dominated
by $\left<B_\perp^2\right>$, which is larger for domain-like than for a
turbulent field. For the domain-like turbulence we have
$\left<B_\perp^2\right>=0.67 B_\mathrm{tot}$. Because of the small variation
in $B_\perp$, the analytical approximation~\eqref{eq:domain-like-approx}
reproduces well the numerical results for the domain-like case. A turbulent
field has, on the other hand, a larger variation in the transverse
magnetic field. Moreover,  the coherence length characterizes only on average
the ``typical''  size of turbulent domains. Both effects lead to a
distance dependence that differs significantly from
Eq.~\eqref{eq:domain-like-approx}.

\section{Direct detection of photon-axion oscillations}
\label{sec:detection}

Although the wiggles after averaging over realisations of the turbulent
field tend to disappear, the oscillatory behaviour in single realisations
may be large. Moreover, the detection of these wiggles is made more
difficult by the finite energy resolution of realistic experiments.
Therefore, it is
common to use either the increased photon survival probability at large
energies (i.e.\ the opacity of the Universe)~\cite{DeAngelis:2007dqd,DeAngelis:2011id, Galanti:2018myb,Long:2021udi,Meyer:2014epa},
the presence of large scale excesses in photon spectra~\cite{Dessert:2021bkv,%
Zhou:2021usu,Mirizzi:2007hr} or the presence of irregularities in photon
spectra (i.e.\ the variance in residuals)~\cite{Wouters:2012qd} to probe the
existence of axion-photon oscillations.

Here we propose instead to use the energy-dependent frequency of the wiggles
themselves as observable. A similar concept
has previously been used to study the Earth-matter effect on neutrino
oscillations~\cite{Dighe:2003jg,Dighe:2003vm,Borriello:2012zc,Liao:2016uis}
by considering the windowed power spectrum
\begin{equation}
  G(k) = |g(k)|^2 = \left|
  \int_{\eta_\mathrm{min}}^{\eta_{\mathrm{max}}}
  \dd{\eta} q(\eta) \me^{\iu \eta k}\right|^2,
  \label{eq:power_spectrum_cont}
\end{equation}
where $\eta$ is a function of energy and $q(\eta)$ is the observed neutrino
flux from a hypothetical source. The oscillations are energy dependent and 
the strongest close to the strong-mixing regime. Thus, the window
$(\eta_\mathrm{min}, \eta_\mathrm{max})$ should be chosen such that it includes
the oscillations that can be resolved, while excluding the strong-mixing
regime to remove noise. For large energies, the oscillation length
$L_\mathrm{osc}\sim 2\pi/\Delta_\mathrm{osc}$ scales as $E$, while at 
low energies it scales with $E^{-1}$. Therefore, we expect clear peaks
in the power spectrum with $\eta\sim E^{-1}$ or $\eta\sim E^{1}$. The windowing
has unfortunately a major drawback that must be handled with care: It induces
a low-frequency peak which may interfere with the signal peaks
(see Eq.~\eqref{eq:power_spectrum_cont} with $q=1$),
\begin{equation}
  G_\mathrm{window}=2\,\frac{1-\cos k\Delta\eta}{k^2}.
  \label{eq:background}
\end{equation}
According to Eq.~\eqref{eq:background}, the power spectrum converges towards
one as $~k^{-2}$.
This method shows the importance of a proper treatment of the magnetic
field:
As discussed in the previous sections, the characteristic signatures
induced by photon-axion oscillations in more realistic magnetic field models are
expected to have a larger cosmic variance and thus to be harder to detect than
in simplified models.

For practical purposes, it is more relevant to consider the discrete power
spectrum
\begin{equation}
  G_N(k)=N\left|\frac{1}{N}\sum_\mathrm{events}\me^{\iu \eta k}\right|^2,
\end{equation}
where the sum goes over detected events. Choosing the correct energy scale
$\eta$ and an optimal window $(\eta_\mathrm{min}, \eta_\mathrm{max})$, one may
observe a peak that exceeds the expected background. 
The location of the signal peak depends on the periodicity of the wiggles and
the chosen window. For example, in a homogeneous magnetic field the photon
survival probability depends on the oscillation length via
Eq.~\eqref{eq:homogeneous_solution}. With $\eta=\Delta_\mathrm{osc}(E)$ the
signal peak will be located around the distance travelled, $k_\mathrm{peak}=s$.
We can, however, look for the generic features discussed
in section~\ref{sec:params}: One expects peaks in two different window
regions and having different energy dependencies, one at low energies with
$\eta\sim E^{-1}$ and one with $\eta\sim E$ at larger energies. 
A similar concept was introduced in Ref.~\cite{Conlon:2018iwn}. There, the idea
was to search for sinusoidal axion signatures in photon spectra by analysing
the Fourier-transformed data or performing a sinusoidal fit.

\begin{figure}
  \includegraphics[width=.5\textwidth]{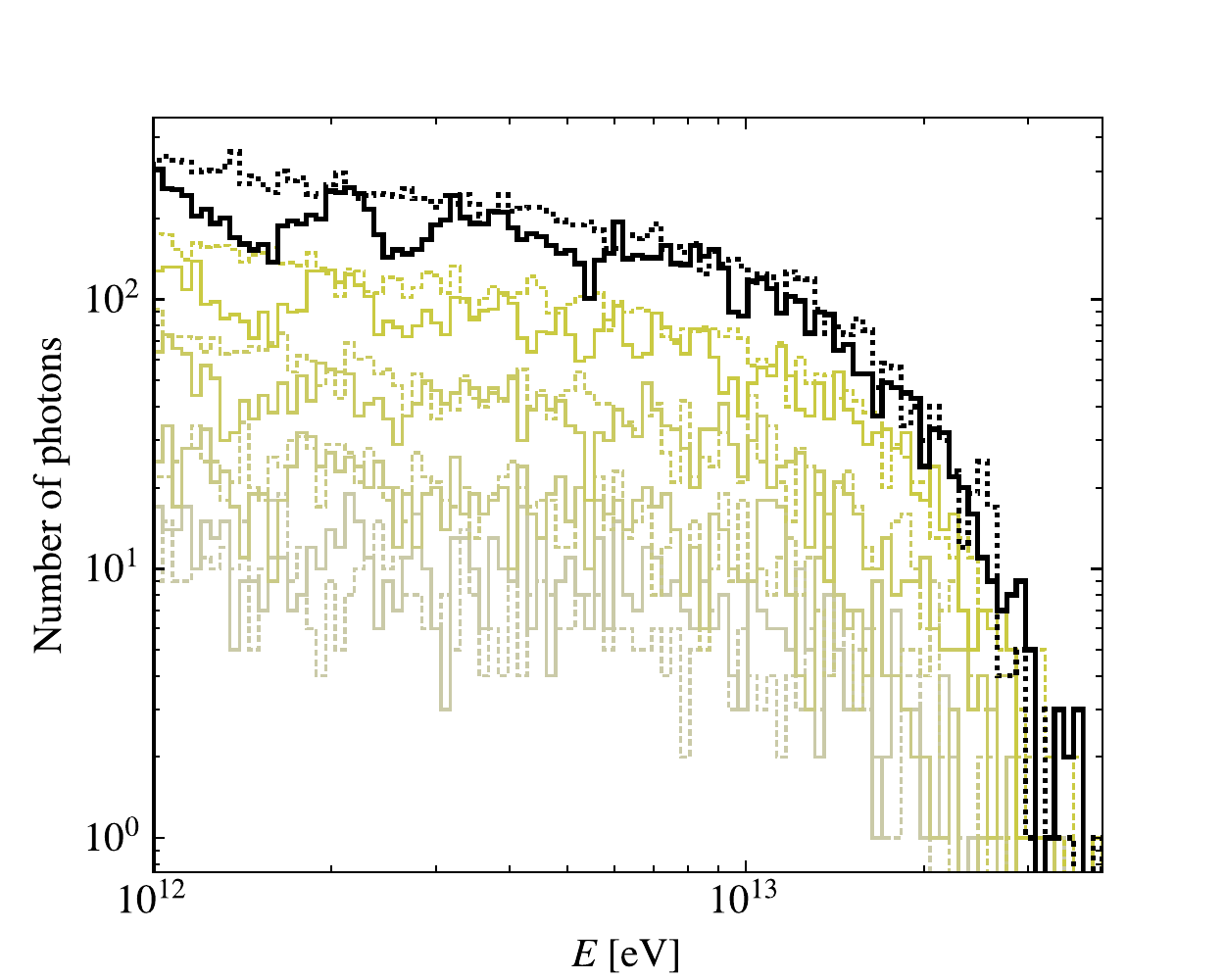}
  \includegraphics[width=.5\textwidth]{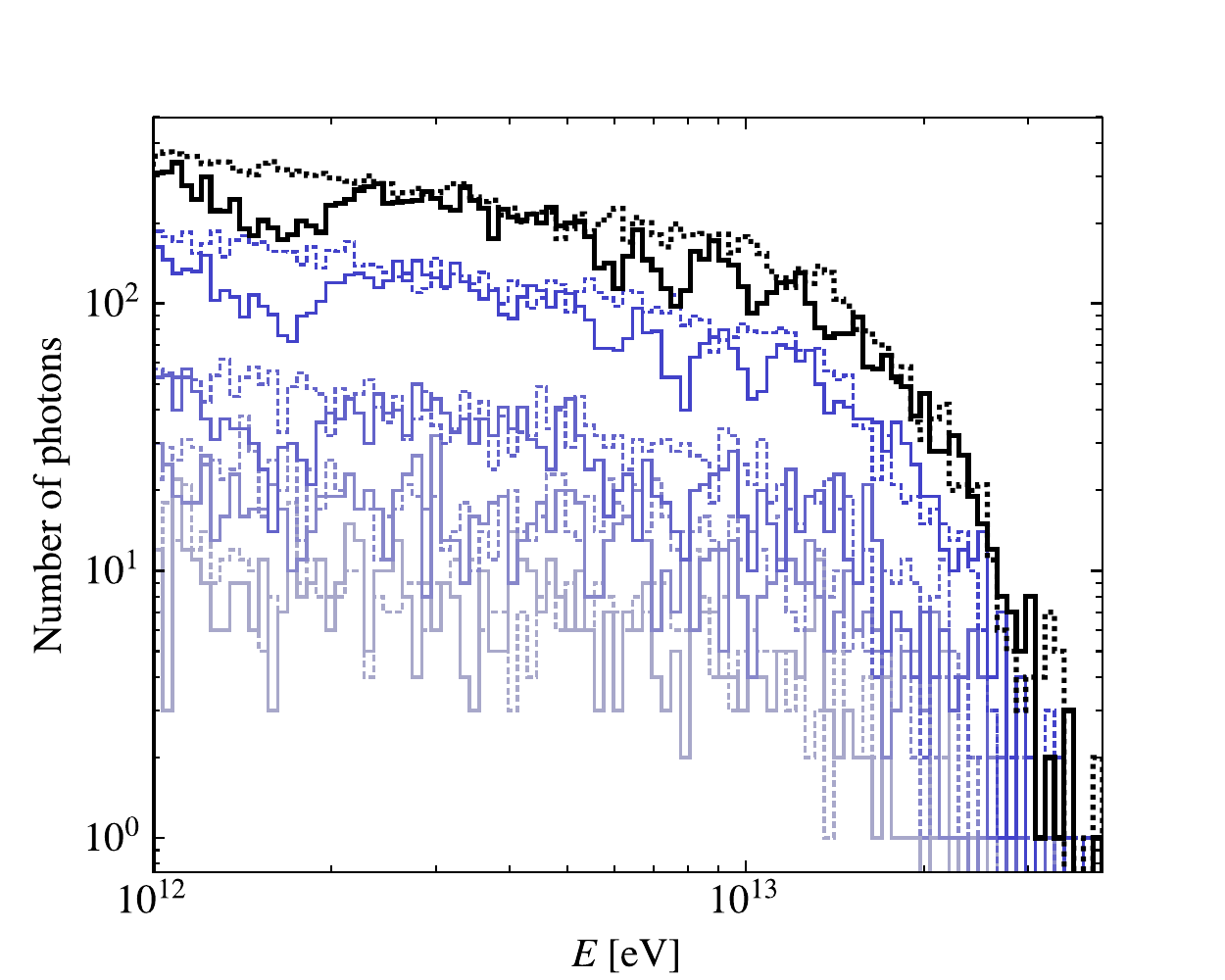}
  \includegraphics[width=.5\textwidth]{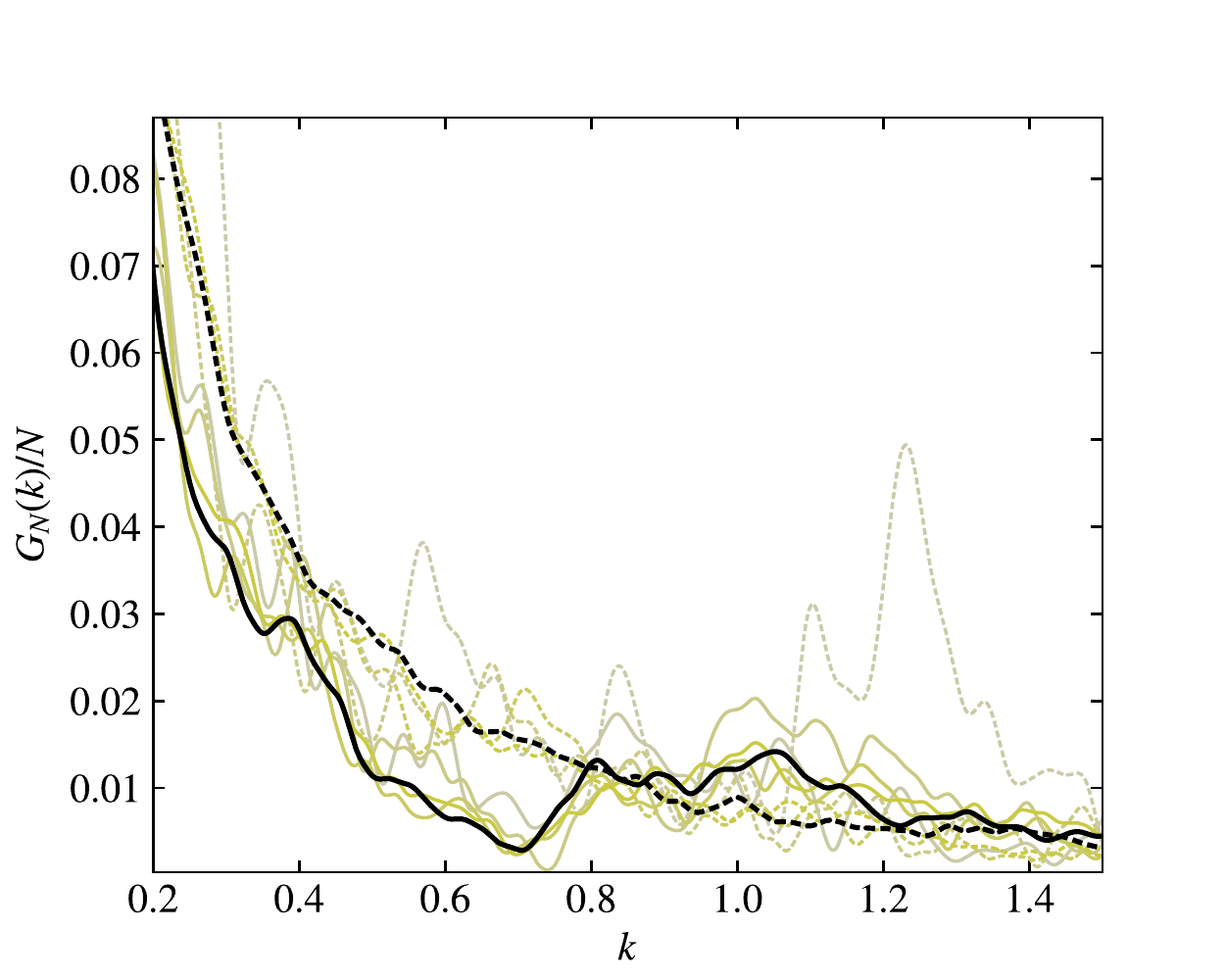}
  \includegraphics[width=.5\textwidth]{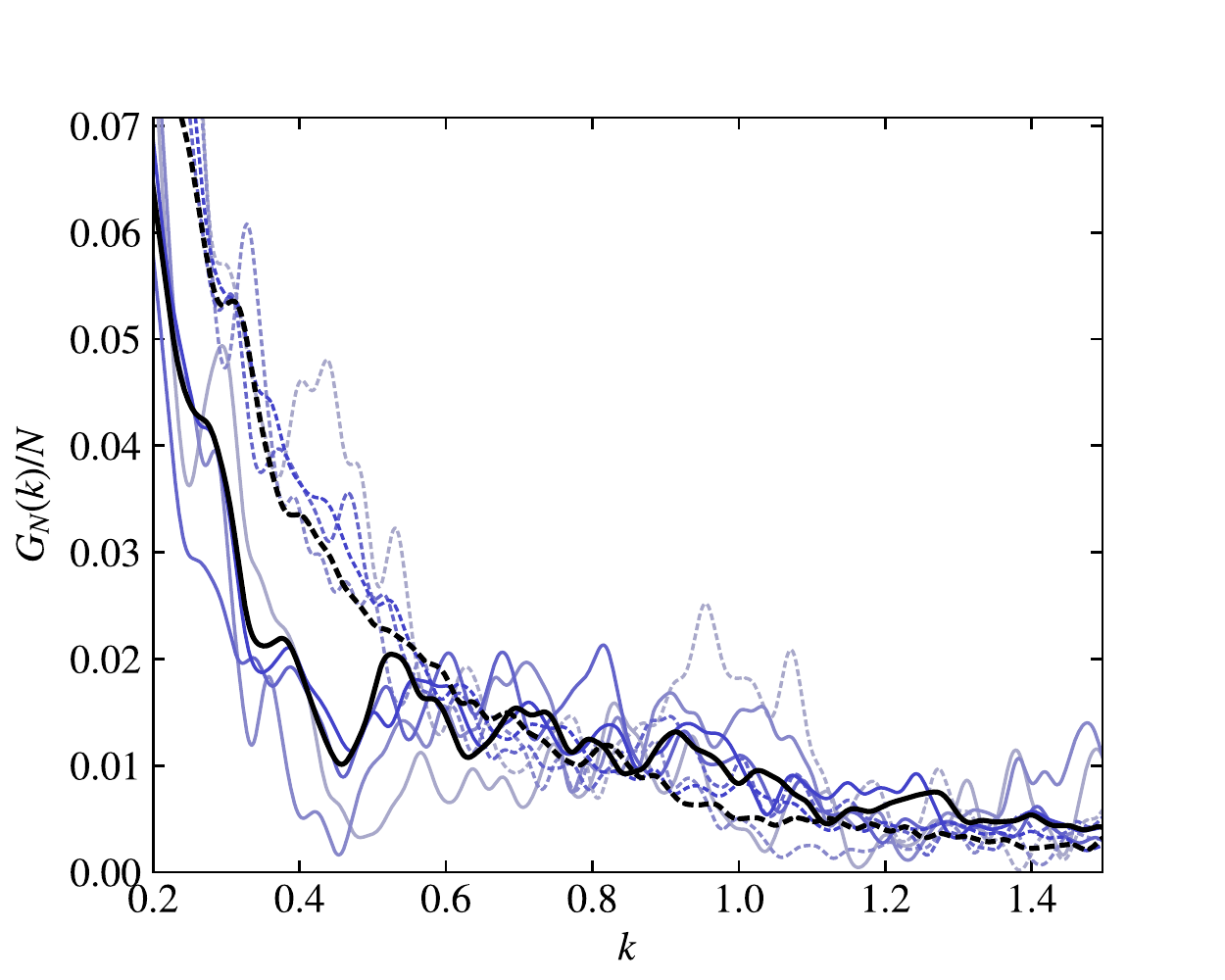}
  \includegraphics[width=.5\textwidth]{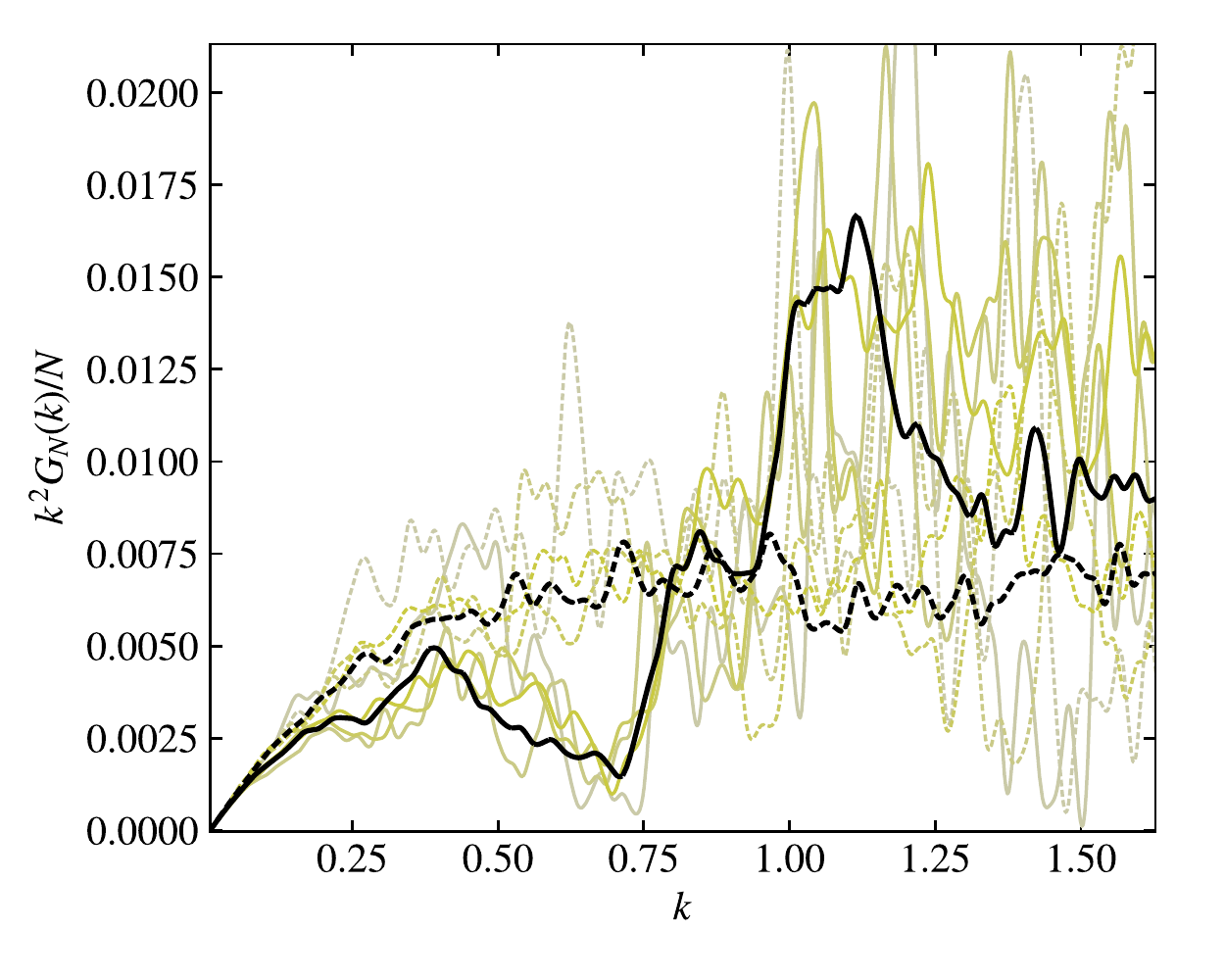}
  \includegraphics[width=.5\textwidth]{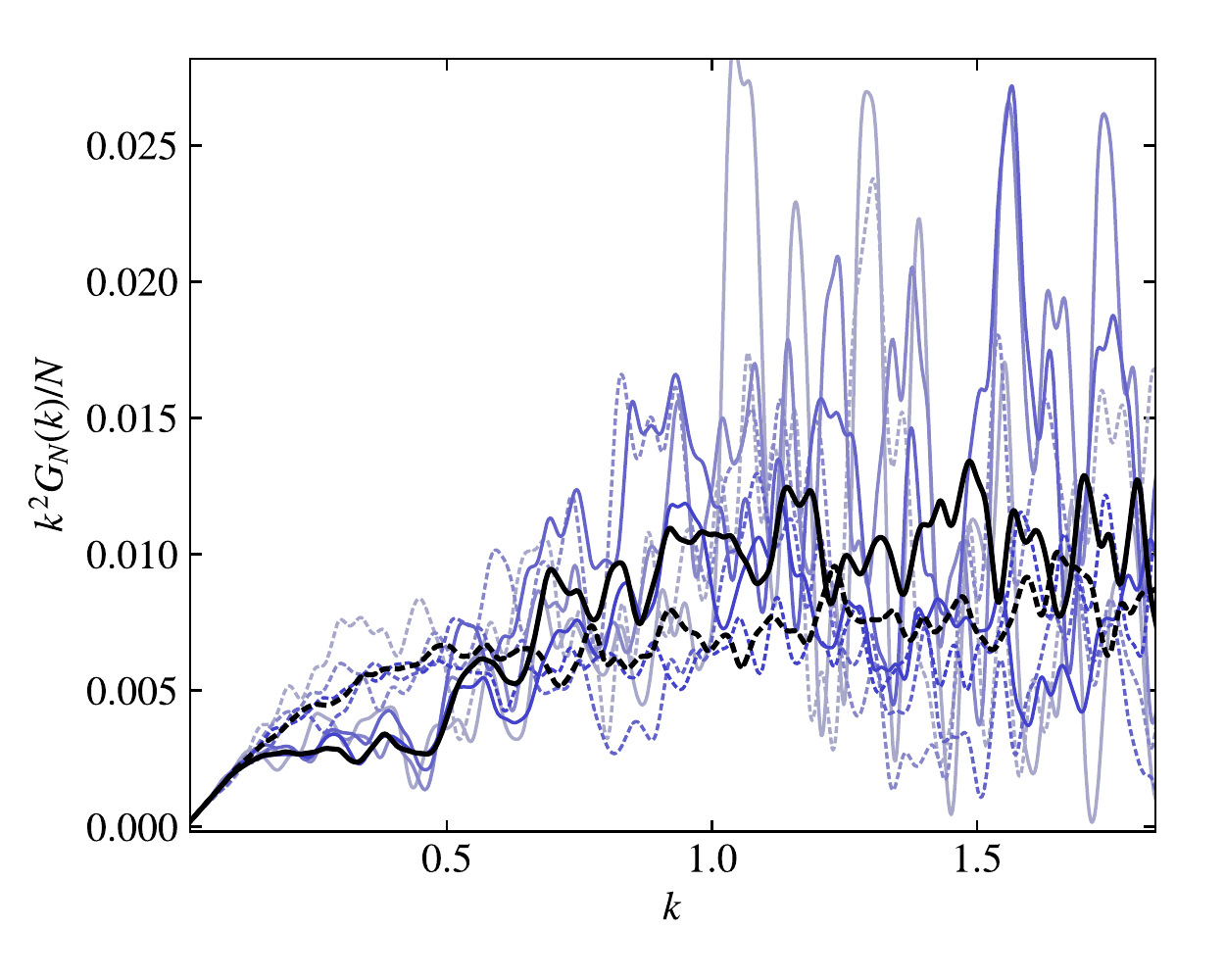}
  \caption{The  photon spectrum with the number of photons per bin (first row),
    photon power spectrum (second row),
  and the photon spectrum multiplied by $k^2$ from a source at redshift
  $z=0.02$ influenced by a Gaussian turbulence are shown for two realisations.
  The parameters in Eq.~(\ref{eq:params}) are used.
  }
  \label{fig:detection}
\end{figure}

To exemplify the suggestion of using the power spectrum to detect photon
axion oscillations, we plot in Fig.~\ref{fig:detection} the  photon
distribution  (first row) and the discrete power spectrum  (second row) for two
realisations of the turbulent magnetic field with the parameters given in
Eq.~(\ref{eq:params}) for a source at redshift $z=0.02$ and a
turbulent field with $B_\mathrm{rms}=5\unit{nG}$ and
$L_\mathrm{max}=10\unit{Mpc}$. In addition, we plot in the last row the
power spectrum multiplied by $k^2$ to better distinguish the peak from the
background. The opaqueness of the lines indicates the number of photons used
in the analysis. The signal peak is clearly visible by eye, and becomes visible
already for $\mathcal{O}(1000)$ photons in these examples. A proper analysis
may therefore yield a significant improvement in the sensitivity compared to
previously suggested approaches. Furthermore, this problem may be a well-suited
for machine learning that potentially can make the method even more sensitive.

Finally, we would like to comment on the interesting
discussions recently given in Ref.~\cite{Marsh:2021ajy}. Here, it was shown
that to first order in the coupling, Eq.~\eqref{eq:eom} can be solved in the
interaction picture outside the strong mixing regime to obtain the
polarised photon-conversion probability\footnote{The integration limit has
been extended to infinity by assuming that $B_i$ vanishes at $s_\mathrm{max}$.}
\begin{equation}
  P_{\gamma_i\to a}(\eta)=\left|-\iu \int_0^{s_\mathrm{max}} \dd{s'}
  \Delta_{a\|}(s')\me^{-\iu \eta s'}\right|^2 =
  \frac{g_{a\gamma}^2}{4}
  \left|\int_0^\infty \dd{s'}B_i(s')\me^{\iu \eta s'}\right|^2
  \equiv \frac{g_{a\gamma}^2}{4} |\mathcal{F}[B(s)](\eta)|^2,
\end{equation}
which in turn can be connected to the auto-correlation function of the
magnetic field. Interestingly, photon-axion oscillations are therefore
(within these assumptions) determined by the auto-correlation function. 
Since the autocorrelation function contains less information than the 
magnetic field, different magnetic fields may share the same 
autocorrelation function. This means that the framework introduced in
Ref.~\cite{Marsh:2021ajy} allows for a more efficient scanning over magnetic
field configurations in axion searches.
There are, however, a few complications that must be 
considered: The description is not valid close to the strong mixing regime
where the oscillations are the strongest, it does not apply to the full energy
range, and one still need to make assumptions on the magnetic fields.
Furthermore, the  photon conversion probability predicted from the
autocorrelation function is highly sensitive to small changes in the
autocorrelation function (see e.g. Fig.~3 in Ref.~\cite{Marsh:2021ajy}).
The most suitable approach to photon-axion oscillations is therefore
in our opinion to look for generic features of
photon-axion oscillations in photon spectra
without considering specific magnetic field environments,
and from that infer information of
the magnetic fields. In fact, computing the power spectrum of the
conversion probability, it follows that
\begin{equation}
  G(k) = \frac{g_{a\gamma}}{4}\int_0^{z_\mathrm{max}}\dd{z} B(z)B^*(z-k)
  =\frac{g_{a\gamma}}{4}\int_{-\infty}^\infty \dd{k} B(k)B^*(k)\me^{\iu\eta k},
\end{equation}
which means that if axions are detected, one can in theory use the power
spectrum of the oscillations, as shown e.g.\ in Fig.~\ref{fig:detection},
to directly infer information about the magnetic field.
One of the main advantages of using the discrete power spectrum
compared to standard approaches of using fit residuals
(such as mentioned in Ref.~\cite{Marsh:2021ajy}), is that no
information on the microstructure of the axion wiggles is lost by the
binning of the data.

\section{Opacity of the Universe}
\label{sec:opacity}

The focus in this work has been on the origin and the characterisation of axion
wiggles in photon spectra, their detection and the effect of the magnetic
fields. However, there is an additional important signature of photon-axion
oscillations that can be used to probe the existence of axions: Since axions
propagate practically without any interactions, there will be an increased
photon survival probability at large energies, thus decreasing the opacity of
the Universe. In Ref.~\cite{Meyer:2014epa}, the difference in the apparent
opacity using a turbulent and a domain-like magnetic field
was already considered.

\begin{figure}
  \centering
  \includegraphics[scale=1]{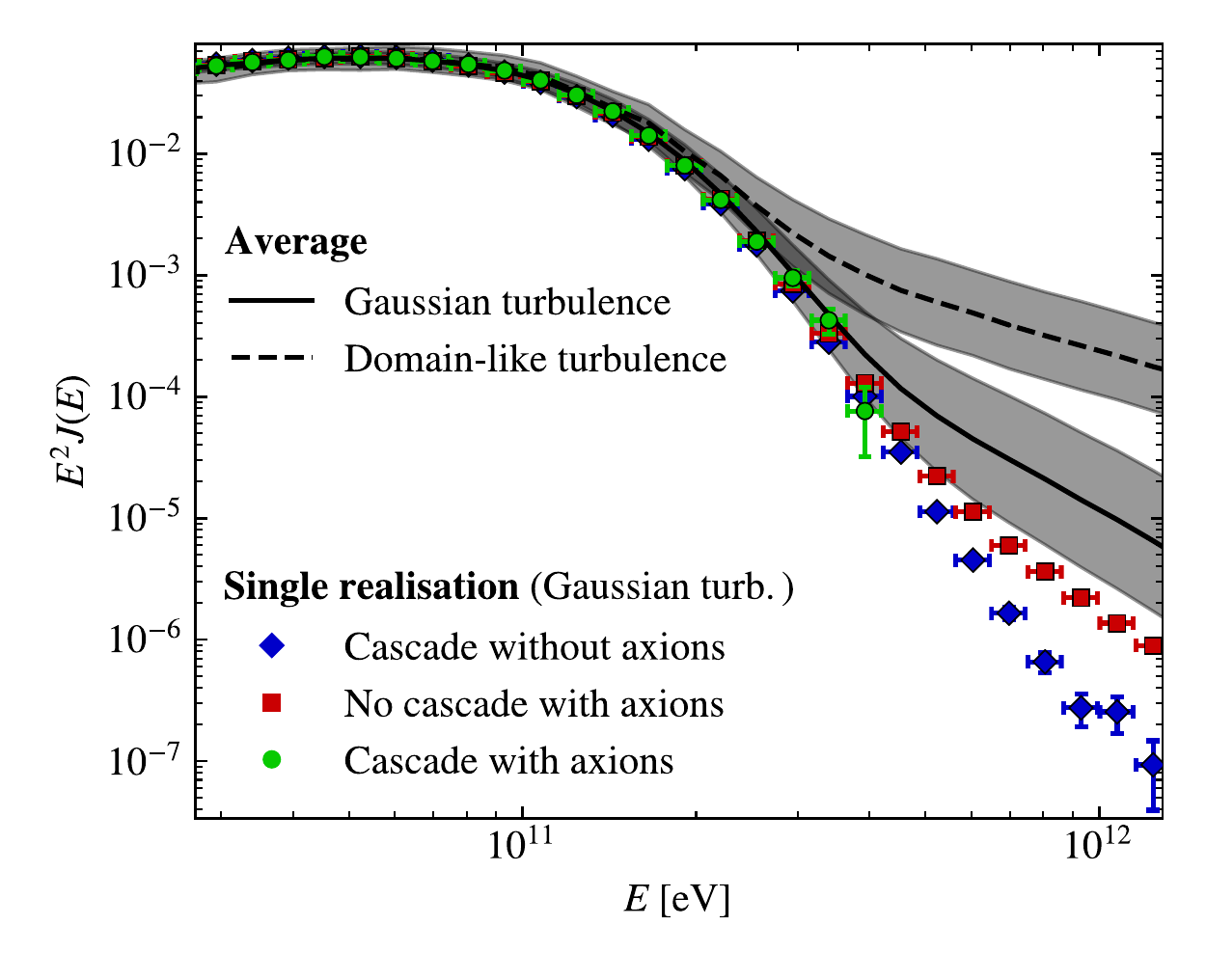}
  \caption{The normalised diffuse photon flux is plotted as a function of
  energy for a source with a power-law index $\alpha=-1.2$ at redshift $z=1$.
  The magnetic field and axion parameters in Eq.~\ref{eq:params} are used.
  The average over many realisations of the magnetic field is shown for
  Gaussian (solid) and domain-like (dashed) as black lines. In order to
  increase the statistics at large energies, the photon absorption is taken
  into account by including an attenuation determined by the photon survival
  probability. Furthermore, the flux obtained for a single realisation is shown
  by red squares, while the flux obtained using the same magnetic field and
  the standard Monte Carlo treatment of the photon absorption as implemented in
  {\tt ELMAG} is indicated by green circles. The errorbars indicate the
  $1\sigma$ Poisson uncertainty. Finally, the flux predicted by {\tt ELMAG}
  without any axions is shown in blue diamonds to visualise the effect that
  the axions have on the flux.
  }
  \label{fig:opacity}
\end{figure}

In order to strengthen our message that one should refrain from using 
domain-like turbulence in quantitative studies on axion oscillations,
we plot in Fig.~\ref{fig:opacity} the
normalised flux from a source with the injection spectrum
$\d N/\d E\propto E^{-1.2}$ at a distance $z=1$ for the magnetic field and
axion
parameters given in Eq.~\eqref{eq:params}. The flux obtained averaged over
many realisations of the magnetic field is shown as a solid (dashed) line for
a Gaussian turbulent (domain-like) field, with the shaded regions corresponding
to  the $1\sigma$ variance between single realisations.
In order to increase the statistics at high energies, we take into account the
photon absorption by including a complex term $\iu \lambda$ in the equation of
motion that describes the mean free path length of the photon\footnote{%
  This implies that only prompt photons are considered. However, the effect
  of cascade photons is negligible since the spectra are dominated by prompt
  photons.},
and update the mean free path length at redshift increments $\Delta z=10^{-3}$.
In other words, the photon attenuation is treated as a continuous change in the
photon survival probability, in contrast to the Monte Carlo treatment of the
electromagnetic cascade that is considered elsewhere in this work. 
As a basis for comparison and to check our numerical calculations, we plot
also the spectrum obtained for a single realisation of the magnetic field using
a Gaussian turbulent field with continuous attenuation (red squares) and
the Monte Carlo treatment of the electromagnetic cascade (green circles).
Furthermore, the spectrum obtained with the electromagnetic cascade without
axions, i.e.\ using {\tt ELMAG} without photon-axion oscillations, is
shown (blue diamonds).
The error bars are computed as the statistical 1$\sigma$ Poisson uncertainty
of the counts in a given bin, and reflect thus only the statistical
uncertainty of the Monte Carlo run\footnote{%
  The number of injected photons in each energy bin
  is uniformly distributed. In the simulation with the continuous attenuation,
  all injected photons contribute to the statistics by a weight corresponding to
  the photon survival probability, making the error bars energy independent.}.

It is clear from Fig.~\ref{fig:opacity} that photon axion oscillations lead
to a decreased opacity of the Universe. However, with the parameters
considered here, the difference between the two treatments of the magnetic
field leads to a significant difference in the predicted  average flux which
is increasing with energy: At $E\simeq 1$\,TeV the difference is around a
factor 20. This result can be understood from
Fig.~\ref{fig:distance_dependence}: The conversion probability 
of photons into axions obtained
using a domain-like field is on average
larger than employing a Gaussian turbulent field at large distances.

The example considered in Fig.~\ref{fig:opacity} further demonstrates the
importance of a proper treatment of magnetic fields in the study of
photon-axion oscillations. While the effect on the opacity on average is
much less prominent for the Gaussian turbulence than for the domain-like
approximation, there is still a significant variation between single
realisations in both cases. The variation is however noticeably larger for
the Gaussian turbulence than for the domain-like approximation.


\section{Conclusion}
\label{sec:conclusion}

In this work, we have studied photon-axion oscillations in a Monte-Carlo
framework based on the {\tt ELMAG} program. We have argued that the use of
statistically averages over magnetic field configurations is misleading
and should be abandoned in the search for signatures from axion-photon
oscillation in the spectra of single sources. Moreover, we have shown
that the predicted signatures---axion wiggles in photon spectra and the
decreased opacity of the Universe---depend strongly on the chosen magnetic field
models. Therefore, over-simplified magnetic field models as the domain-like
field should be used with care for quantitative predictions.

We have discussed mainly those characteristic signatures of axion-photon
oscillations which are independent of the concrete astrophysical
environment. In particular, the oscillation length will scale as $E^{1}$
below the threshold energy $E_\mathrm{min}$ and as $E^{-1}$ above the
threshold energy $E_\mathrm{max}$, while it will be constant at intermediate
energies. We have proposed to use this energy dependence as an observable
in the search for photon-axion oscillations using the discrete power
spectrum. This method can in principle be also used to infer information about
the magnetic field environment from the observation of axion wiggles.


\section*{Acknowledgments}
We would like to thank Alessandro Mirizzi for providing valuable comments on the
manuscript.



\newpage
\appendix

\section{Varying axion mass}

\begin{figure}[bht]
\centering
\includegraphics[scale=1]{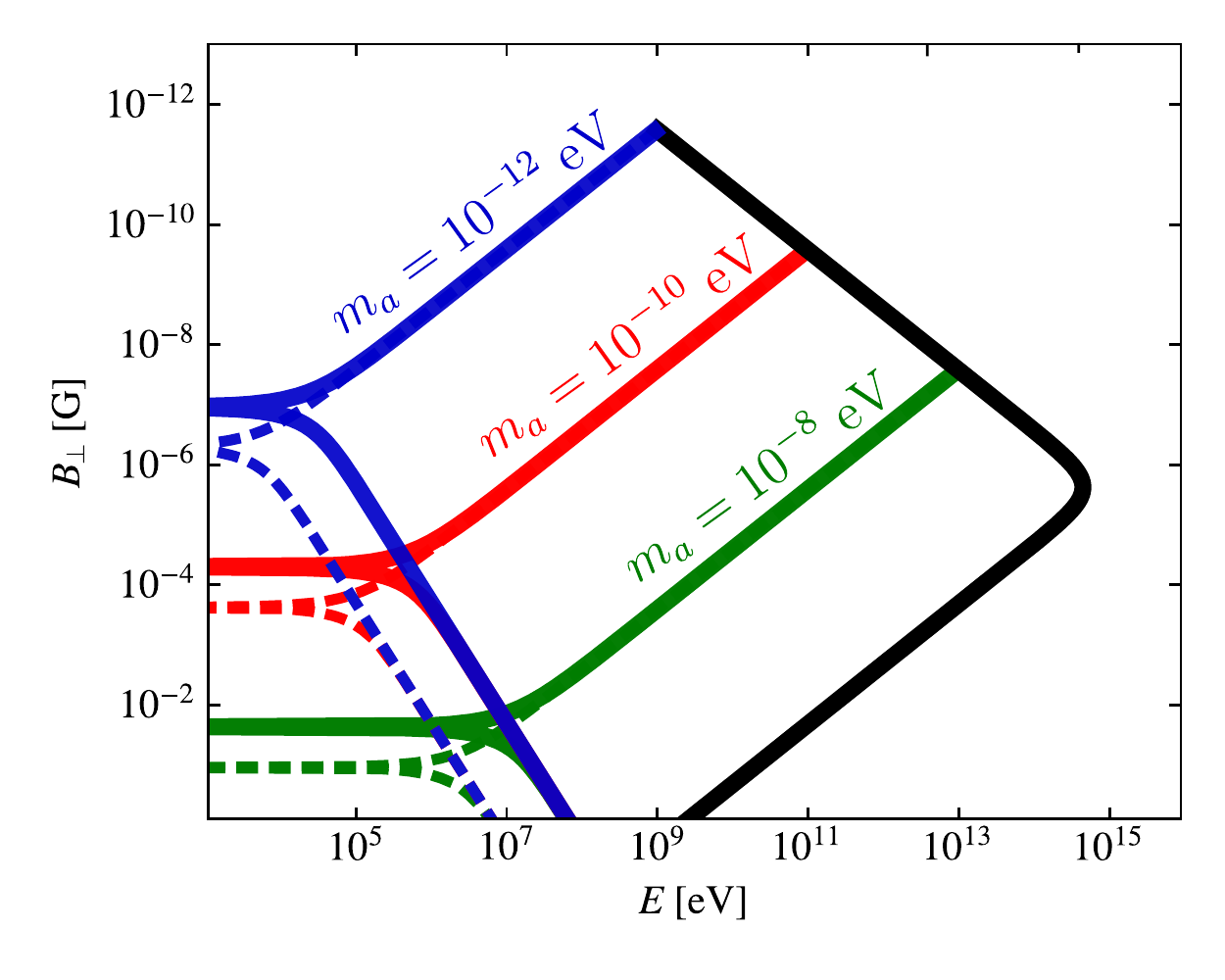}
\caption{
  The transition energies in Fig.~\ref{fig:param2} is replotted for different
  axion masses. In addition, the effect of
  decreasing the reference electron density to  $n_{e,0}=0.002\unit{cm^{-3}}$
  is shown by dashed lines.
}
\label{fig:param3}
\end{figure}

For completeness, we plot in Fig.~\ref{fig:param3} the transition energies from
Fig.~\ref{fig:param2} varying the axion mass and plasma density.
The high energy transition to the strong mixing regime (black line) is
independent of the axion mass and plasma density, while the low energy
transition is decreasing for a decreasing axion mass and plasma
density. Interestingly,
this means that a weak magnetic field only will produce axion wiggles in photon
spectra if the axion mass is sufficiently small. For example, if
$m_a\sim 10^{-12}\unit{eV}$, the magnetic field strength must be stronger 
than $\sim10^{-11}\unit{G}$ to produce wiggles. On the other hand, if
$m_a\gtrsim 10^{-6}\unit{eV}$, the strong-mixing regime disappears and
therefore no significant axion wiggles are produced.

\end{document}